\documentclass[review]{elsarticle}

\usepackage{amssymb}
\usepackage{subfigure}
\usepackage{color}
\usepackage{amsthm}
\usepackage{mathtools}
\usepackage{lineno,hyperref}

\journal{Journal of \LaTeX\ Templates}

\begin{document}

\begin{frontmatter}


\author[First]{G Kamalakshi }
\ead{ranikamalakshi@gmail.com}

\author[First]{Prita Pant}
\ead{pritapant@iitb.ac.in}

\author[First]{M P Gururajan \corref{cor1}}
\ead{guru.mp@iitb.ac.in,gururajan.mp@gmail.com}

\cortext[cor1]{Corresponding author.}

\address[First]{Department of Metallurgical Engineering and Materials Science, Indian Institute of Technology Bombay, 400076, Mumbai, Maharashtra, India.}

\title{Stacking fault energy induced softening in the nucleation limited plasticity regime: a molecular dynamics study on Cu-Al alloys}


%

\begin{abstract}
In general, with the addition of solutes, the yield strength of alloys is expected to increase and this phenomenon is known
as solid solution strengthening. However, reports on the ``anomalous'' softening with alloying additions are not uncommon; 
for example, it is known that the heterogeneous nucleation of dislocations at the sites of solute atoms can lead to softening.
In this study, using Molecular Dynamics (MD) simulations, we show anomalous softening in Cu-Al alloys deformed at 300 K; 
specifically, not only the yield stress but also the magnitude of stress drop at yield decrease with increasing Al content.
We calculate the stress needed for the homogeneous nucleation of partial dislocation loops using a 
continuum model. One of the key inputs to the continuum model is the stacking fault energy (SFE). We carry out the  
thermodynamic integration 
to evaluate the free energies in crystals with and without the stacking faults and hence calculate the SFE 
as a function of Al 
content at 300 K. Using the SFE values thus obtained we show that softening is a consequence of the reduction of SFE values with the addition of Al in a system where yielding is controlled by the nucleation of dislocation loops. Specifically, the results of continuum model are in good agreement with MD simulation results of pure 
copper. However, in the Cu-Al alloys, the drop in yield strength can only be explained using a combination
of the continuum model (which assumes homogeneous nucleation), and the reduction in heterogeneous nucleation 
barrier (which is a function of the ratio of the unstable and stable stacking fault energies). We show that 
the decrease in stress 
drop can be rationalised in terms of the stored  energy available at yielding -- the stored energy at 
yielding decreases with 
increasing Al, and, as a result, the maximum dislocation density decreases with increasing Al content. 

\end{abstract}




\end{frontmatter}


\section{Introduction}

Conventionally, the addition of a solute is expected to increase the yield stress in alloys~\cite{Dieter,Courtney,MayersChawla}. 
However, solid solution softening is not uncommon; for example, it has been reported in Molecular Dynamics (MD) simulations of 
Cu-Pb system by Rupert~\cite{TJRupert}, Cu-Ag system by Amigo et al.~\cite{CuAg}, and Cu-Sb system by Rajgarhia et al.~\cite{RKRajgarhia}. 
Rupert reports that in nanocrystalline materials yield strength is proportional to the Young's modulus (similar to the metallic glasses); 
and, hence attributes the softening behaviour to the reduction in Young's modulus due to the addition of Pb to Cu. Rajgarhia et 
al.~\cite{RKRajgarhia} correlate the softening to unstable stacking fault energy near the solute atoms, which is related
to the energy barrier for the Shockley partial dislocation nucleation. Similarly, in Cu-Ag~\cite{CuAg} system also, the reasons for 
decreasing yield stress is discussed with respect to the change in unstable stacking fault energy near Ag atoms. Thus, in both the 
systems (Cu-Ag and Cu-Sb), the softening is a result of the addition of solutes, which decrease the unstable stacking fault energy 
near the solute atoms, which, in turn reduce the stresses required for dislocation nucleation. Thus, the reduction in the energy occur 
only if solute atoms are present at the fault plane and the dislocation nucleation is heterogeneous. 

Interestingly, it is known that decreasing stacking fault energy (SFE) can lead to solid solution softening. For example, 
studies in which the stacking fault energy (SFE) of copper and aluminium was changed by using different interatomic potentials, 
as the SFE is decreased, the deformation is dominated by partials~\cite{DiffSFECu,DiffSFEAl}; specifically, in the case of copper 
bi-crystals, the same trend, namely decrease in yield stress with decreasing SFE is observed. This leads to the question
as to whether alloying additions which lead to a decrease in SFE can lead to solid solution softening. In this paper, 
using Cu-Al as the model system, we show that alloying additions that lead to decrease in SFE do indeed lead to solid solution 
softening -- even when the dislocation nucleation is homogeneous.

 The Cu-Al system (at the copper-rich end of the phase 
 diagram) is an ideal system to study the effect of stacking fault energy on deformation behaviour. This is because,
\begin{itemize}
\item Experimentally, it is known that the addition of Al to Cu lowers the SFE values from 
 78 mJ/m$^2$~\cite{MayersChawla,Rohatgi}, or, 41 $\pm$ 9 mJ/m$^2$~\cite{CarterRay} for pure copper 
 to about 6 mJ/m$^2$ for nearly 13 at.\% Al alloy~\cite{Rohatgi}; and, 
\item From the phase diagram, it is known that the alloy remains as a single phase material (of face centered cubic
crystal structure) in this chosen range of composition~\cite{phasedia}. 
 \end{itemize}
Thus, in this study we have chosen the Cu-Al system for our molecular dynamics studies on deformation behaviour at 300 K as a function
of Al content (and, hence, as a function of decreasing SFE). 

At this point, we want to emphasise that the experimental studies on Cu-Al alloys~\cite{Rohatgi,Yin2016,Liu2015a,Szczerba2017} 
show solid solution strengthening i.e  the strength of the material increases with increasing the Al content. This, we believe, is because, 
experiments are typically done on polycrystalline materials containing both  dislocations and grain boundaries; the latter  can act as both 
source and sink for dislocations. In contrast, our simulations are carried out in the regime of dislocation nucleation limited plasticity.
On the other hand, it is known that in deformation experiments on small scale systems (such as micro-pillars, nano wires and nano whiskers) yielding can be dominated
by dislocation nucleation~\cite{Bei2007,lu2011surface,Gunther2009,Chen2015}). Hence, the current simulation study indicates that if experiments are carried out on Cu-Al system using 
micro-pillars or nano-whiskers, we might expect solid solution softening as a function of increasing Al content. 

The rest of this paper is organised as follows: in Section~\ref{Section2}, we begin with a description of simulation methodology;
this includes a description of our tensile deformation simulations, the choice of interatomic potentials and the determination of 
appropriate system size for the simulations; specifically, the evaluation of lattice parameters, and elastic moduli are used
to decide on the choice of interatomic potentials. In order to unambiguously rationalise the observed deformation mechanisms in different
Cu-Al alloys, it is essential to know the SFE that results from the interatomic potentials used by us. Using
Frenkel-Ladd technique, we carry out the thermodynamic integration to evaluate the free energies of systems with and without
stacking faults -- for various compositions of Cu-Al alloys at 300 K (at which the deformation simulations are carried out)
-- and, evaluate the SFE from the resulting free energy values. We use the SFE thus evaluated for all further discussions 
and interpretation of the simulation results. Thus, the thermodynamic integration is an important component of our study.
In Section~\ref{Section3}, we describe the thermodynamic integration as well as the evaluation of generalised stacking fault energy 
(GSFE) curves. In Section~\ref{Section4}, we present the results of tensile deformation simulations in Cu and Cu-Al alloys. 
We correlate the deformation behaviour (in terms of the yield point and the drop in stress at the yield point) with the defect 
and micro- structures. In Section~\ref{Section5}, we discuss the results of solid solution softening; further, we correlate the 
softening to nucleation of partials using a combination of a continuum model for homogeneous nucleation of partials (for which, all the inputs are taken from the MD simulations
themselves -- using the same interatomic potentials) and a model for reduction in heterogeneous
nucleation barrier as a function of the ratio of unstable and stable stacking fault energies. We conclude the paper with a summary of salient results in 
Section~\ref{Section6}.

\section{Simulation methodology} \label{Section2}

We have performed Molecular Dynamics (MD) simulations of tensile deformation on pure copper and Cu-Al alloys (of compositions
4.6, 8.9 and 13.0 at.\% Al) at 300 K. We consider single crystals having the face-centred-cubic crystal structure. For carrying out the molecular dynamics simulations on a high performance 
computing cluster,  we have used LAMMPS~\cite{lammps}. The microstructures are analysed using Open Visualisation  Tool 
(OVITO)~\cite{ovito}. 

\subsection{Simulation set-up}
 
\begin{figure}
\begin{center}
\includegraphics[scale=0.2]{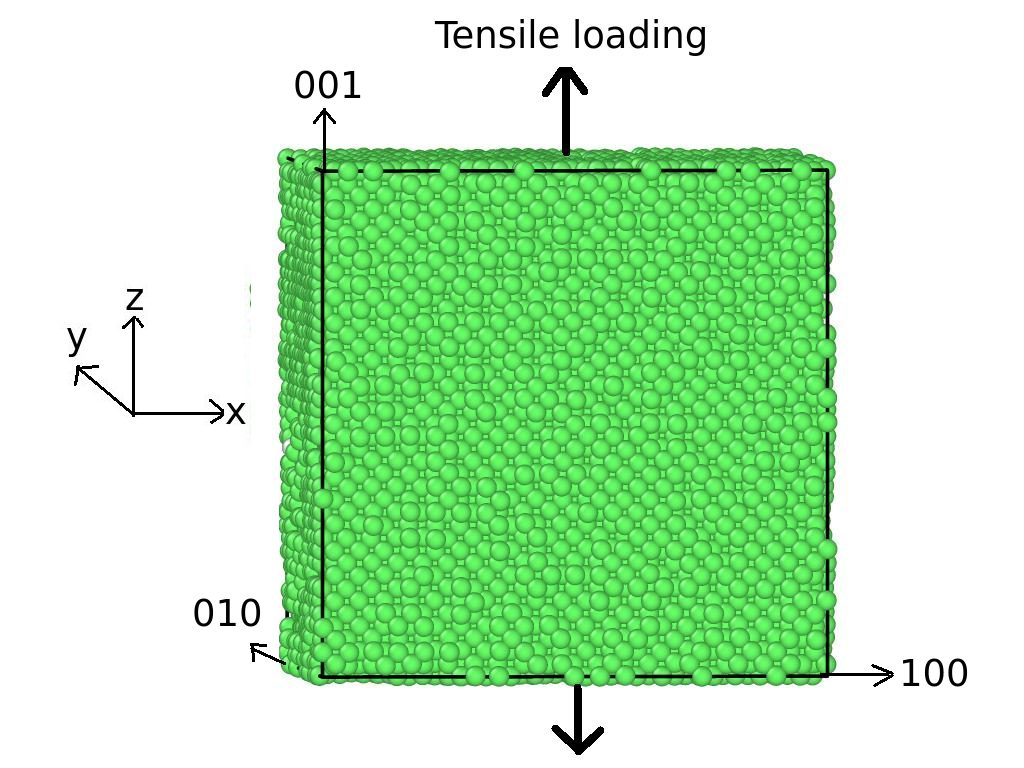}
\includegraphics[scale=0.2]{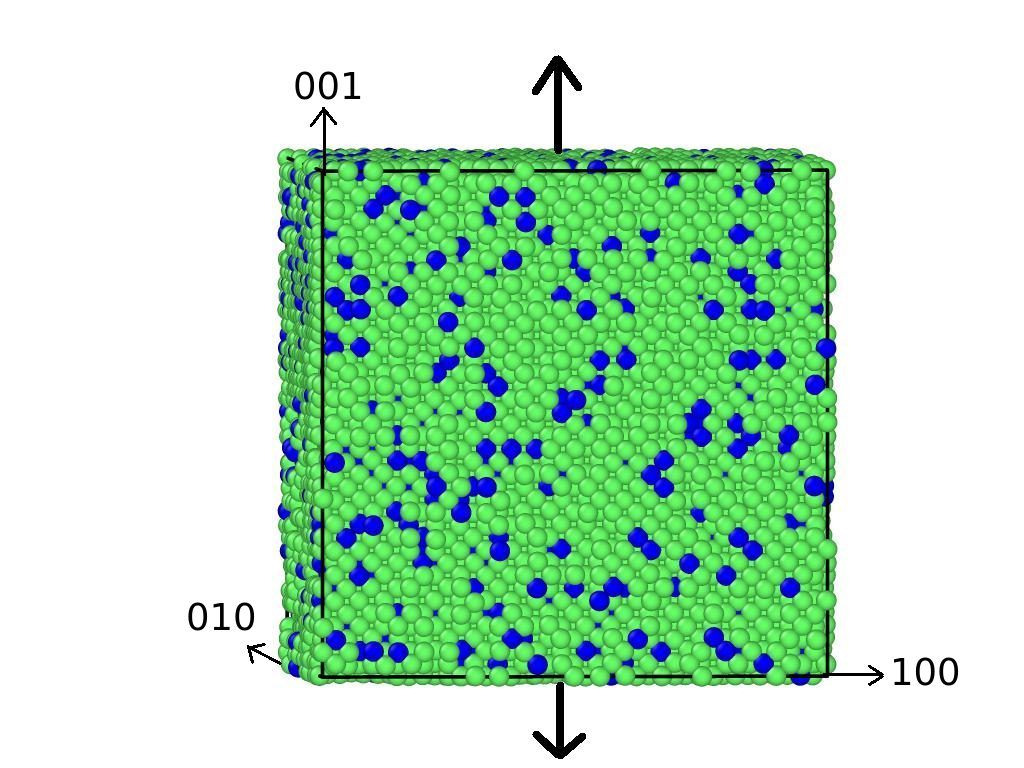}
\caption{Schematic of simulation set-up for uniaxial tensile loading along z axis for (a) Cu and (b) Cu-13 at.\%Al. Green circles represent Cu atoms and the blue circles represent Al atoms.}\label{SchematicLaoding}
\end{center}
\end{figure}

The [100], [010] and  [001] crystallographic directions of the Cu/Cu-Al alloys are aligned along the 
x-, y- and z-axes of the simulation cell. 
Uniaxial tensile loading is done along the z-axis at a strain rate of $10^{8}$ $s^{-1}$ as indicated in the Figure.~\ref{SchematicLaoding}. The simulations are done under periodic 
boundary conditions 
along all three directions; this helps us avoid surface effects. The tensile simulations are performed at 300 K and 0 Pa using an NPT 
ensemble on
systems of size 30 unit cells $\times$ 30 unit cells $\times$ 30 unit cells. The temperature and pressure are maintained using 
Nos\'{e}-Hoover thermo- and baro-stats respectively. For each composition, three simulations are performed. 
The reported values of Young's modulus, yield stress and 
the drop in stress at the yield point are the averages obtained from the stress-strain plots of these three simulations; 
the standard deviations are used to indicate the error in the data.

The initial configuration for the simulations are prepared as follows: at 0 K, systems with different lattice 
parameters are set up and the energy of the system 
is minimised using conjugate gradient method. This leads to the equilibrium lattice parameter at 0 K; we have confirmed that the 
equilibrium lattice parameter thus reached is independent of the initial choice of lattice parameter. Then, the velocity is 
rescaled to correspond to 100 K. The temperature is then ramped from 100 K to 300 K in 100 ps using an NPT (isothermal-isobaric) 
ensemble.  Relaxation is done for 100 ps in an NPT  ensemble at 300 K, so that the system pressure, temperature and lattice 
parameter values are stabilized before loading. We have confirmed that relaxation preformed for 
longer times ( $>$ 100 ps) do not change the results. In the case of Cu-Al alloys, the initial configuration also involves replacing 
a fraction of Cu atoms by Al atoms to obtain the required composition.  These Al atoms are randomly distributed to obtain different 
initial configurations for any given composition.

\subsection{Choice of the interatomic potentials}

One of the key ingredients for the MD simulations is the interatomic potentials; the embedded atom method (EAM) potential 
for elemental metallic systems as well as for binary and ternary alloy systems are available: see the NIST potential
site~\cite{NISTPotential} and LAMMPS potential folder, for example. 

Zhou et al., have developed
a model which allows for potentials for binary systems developed based on elemental potentials~\cite{Zhou2004}. We denote
the EAM potential that we have used for (Cu-rich) Cu-Al alloys based on the Zhou et al., model as PZ.  Ward et al.~\cite{ZW} 
have developed a computationally less expensive method to create the binary potentials using elemental Embedded Atom Method (EAM) 
potentials. They report that the material properties computed using the  potentials developed using their model agrees well 
with the existing literature; and that the potentials developed using their  model is superior to the alloy potentials developed 
using the model of Zhou et al. We denote the EAM potential that we have used for (Cu-rich) Cu-Al alloys based on the model
of Ward et al. (using the elemental potentials reported by Zhou et al), PZW. In this subsection, we report the lattice
parameters and elastic constants generated using these two potentials in order to benchmark these potentials. 

\subsubsection{Lattice parameters}

\begin{figure}[htpb]
\centering
\includegraphics[trim=0.5cm 0.5cm 0.5cm 0.5cm,height=3in,width=3in]{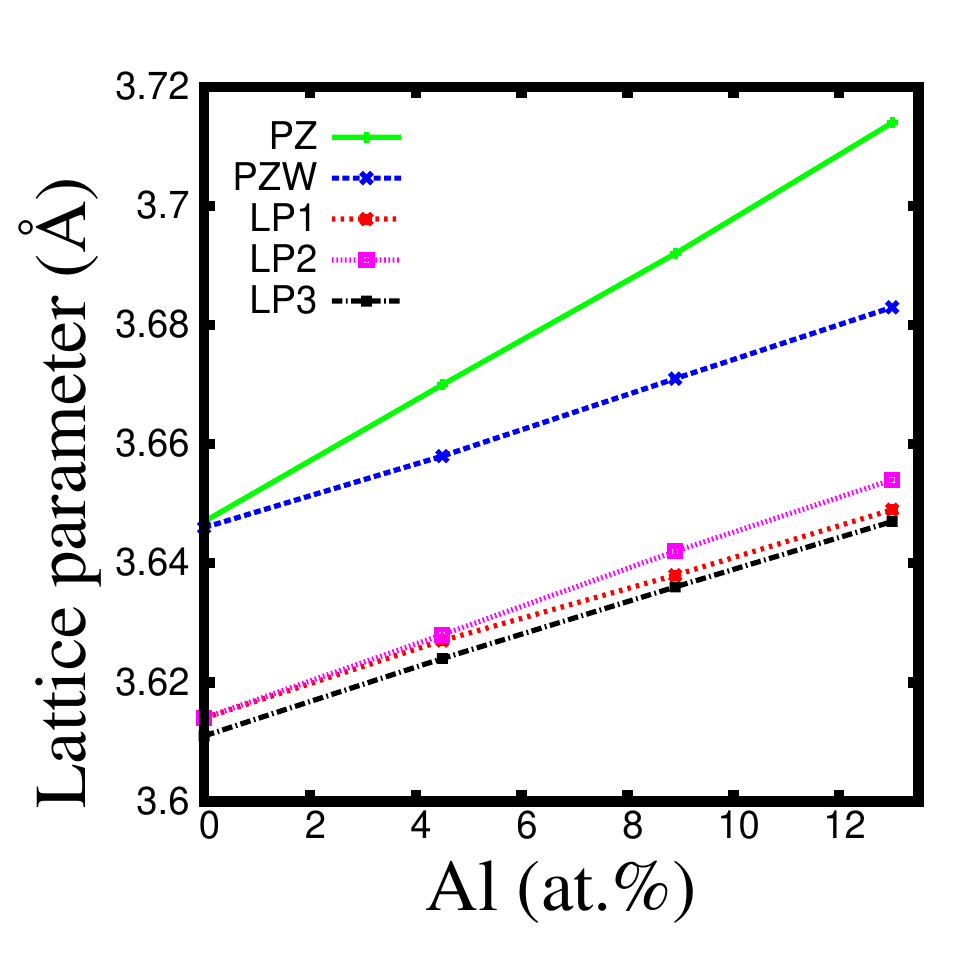} 
\caption{Variation of lattice parameter with Al composition. Here LP1, LP2 and LP3 indicate the 
lattice parameter values reported by Cain et al.~\citep{CainThomas}, Seshadri et al.~\cite{Seshadri1979}, and Tomokiyo et al.~\cite{Tomokiyo1987}, respectively.}\label{LPall}
\end{figure}

The lattice parameter values of Cu and Cu-Al alloys are calculated as follows: initial simulation setup is done as explained in the previous section. After reaching the required temperature of 300 K, relaxation is done for 100 ps in an NPT  ensemble at 300 K. The lattice parameter is calculated by dividing the average box length along any of the three crystallographic directions by the number of unit cells along that direction. The lattice parameter values thus obtained using these two potentials 
are shown in Fig.~\ref{LPall} along with experimental values. We have fit the simulation as well as experimental data to a stright line.
From the slope of the straight line fit, we obtain the coefficient of increase of lattice parameter with composition (Vegard's 
coefficient); the Vegard's coefficient obtained for PZ potential is 0.0051 and that for the PZW potential is 0.0028 while
those obtained from the experimental data reported by Cain et al.~\citep{CainThomas}, Seshadri et al.~\cite{Seshadri1979} and Tomokiyo et al~\cite{Tomokiyo1987} are 0.0026, 0.0031 and 0.0027 respectively. As is clear from the figure, and the Vegard's coefficients, we can conclude that the rate of change of lattice parameter with Al addition is large in the case of PZ potential and that of PZW potential is in good agreement with the experimentally reported values.

%
%

\subsubsection{Elastic constants}

Copper is a cubic material and hence there are three independent elastic moduli, namely, $C_{11}$, $C_{12}$ and  $C_{44}$~\cite{JFNye}.
Using both the PZ and PZW potentials we have evaluated these constants.

For obtaining the elastic constant tensor, we have followed the methodology indicated in the solved example~\cite{EConstant} 
of LAMMPS: The simulation box is strained by $10^{-6} $along the x-direction (Voigt deformation component 1). The energy minimization 
of the system after straining gives the pressure tensor hence the elastic constants 
$C_{11}$, $C_{21}$, $C_{31}$, $C_{41}$, $C_{51}$ and $C_{61}$ can be calculated. The complete elastic constant tensor is obtained 
by repeating the procedure for other directions. For cubic system the only non-zero independent 
components are $C_{11}$, $C_{12}$ and  
$C_{44}$. 

\begin{figure}
\begin{center}
\includegraphics[scale=0.7]{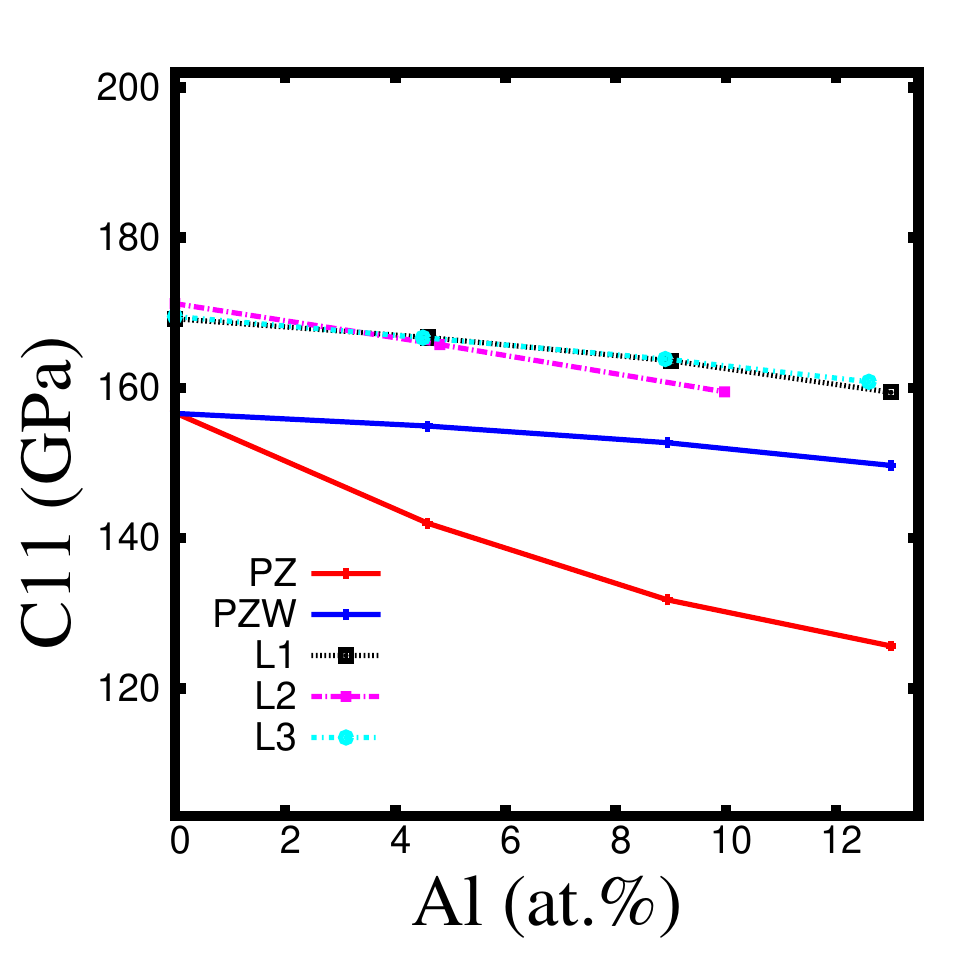}
\includegraphics[scale=0.7]{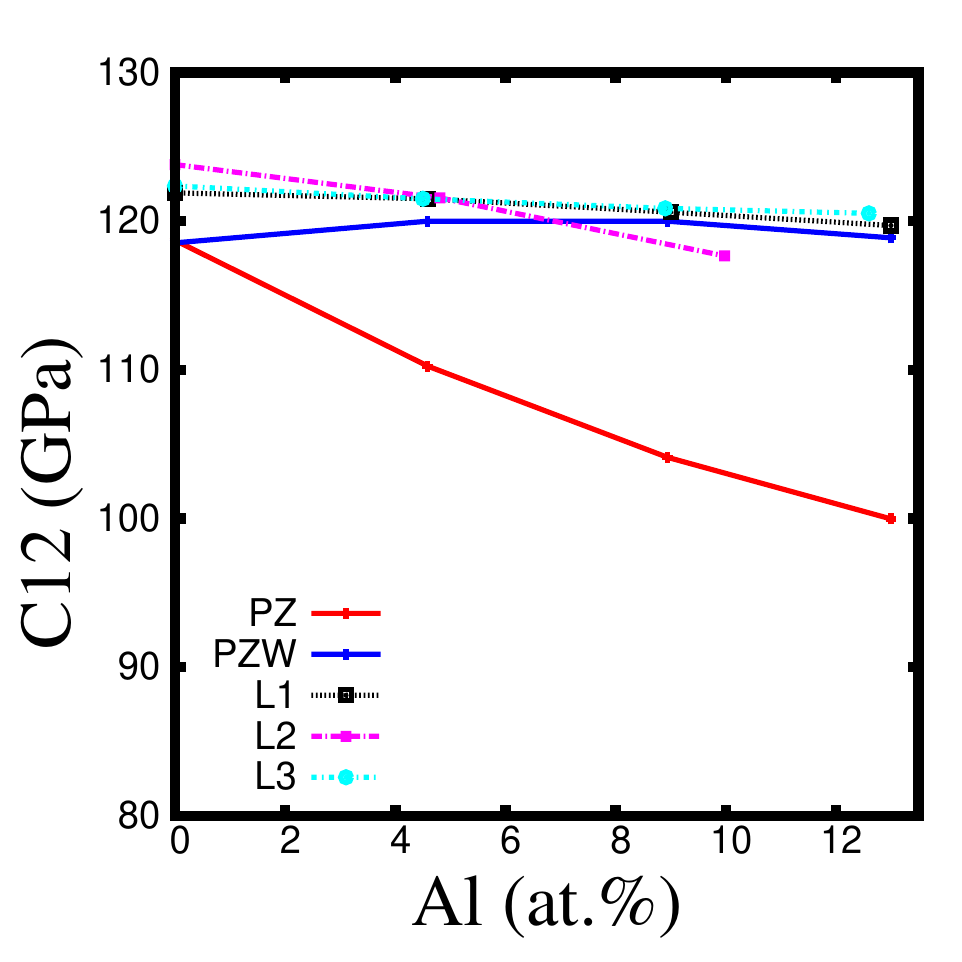}
\includegraphics[scale=0.7]{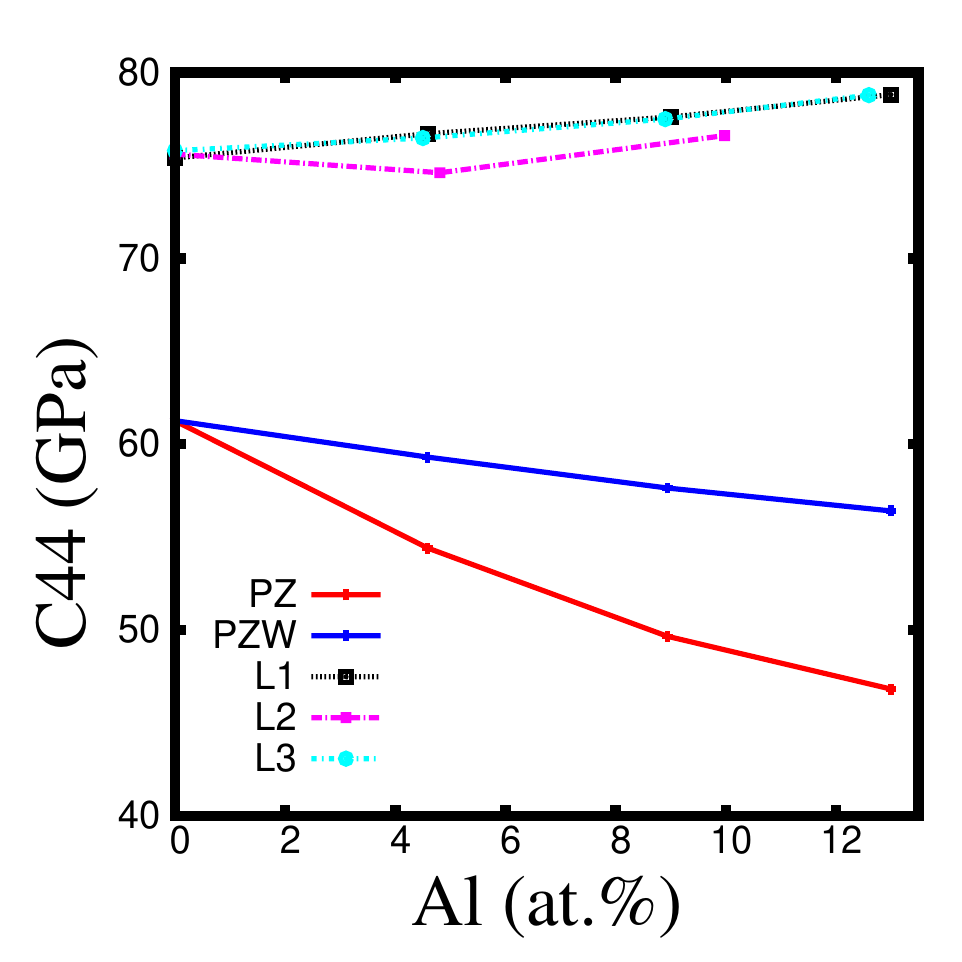}
\caption{Variation of (a) $C_{11}$, (b) $C_{12}$ and (c) $C_{44}$ with Al composition. Here L1, L2 and L3 indicate the moduli 
values reported by Cain et al., Neighbours et al. and Moment respectively.}\label{ElasticCons}
\end{center}
\end{figure}

In Fig.~\ref{ElasticCons}, the values of $C_{11}$, $C_{12}$ and  $C_{44}$ values obtained using both the PZ and PZW potentials
are shown as a function of Al composition. In the same figure, we have also indicated the experimental data reported by 
Moment~\cite{Moment1972}, Neighbours et al.~\cite{Neighbourst} and Cain et al.~\cite{CainThomas}. While $C_{11}$, $C_{12}$ 
and  $C_{44}$ broadly decrease with increasing Al content for both the potentials,
the slopes are not the same. Further, the changes with Al addition are not monotonic;
for example, $C_{12}$ values first increase slightly and then decrease with increasing Al content. 
 Further, note that in the case of experimental data, the (overall) trend for $C_{44}$ (increasing 
with increasing Al content) is not the same as the trend for $C_{11}$ and $C_{12}$ (decreasing with increasing Al content); 
this is considered anomalous~\cite{Moment1972}. In addition,
in the experiments of Neighbours et al., the change of $C_{44}$ with Al composition is not monotonic.

Using the values of $C_{11}$, $C_{12}$ and  $C_{44}$, we can calculate the Young's modulus as $E_{th}$ = $1/S_{11}$ ( where $S_{11} = 
\frac{C_{11}+C_{12}}{(C_{11}-C_{12})(C_{11}+2C_{12})}$~\cite{JFNye}). We have also compared the Young's modulus thus obtained
with the slope of the initial linear portion of the stress-strain curve $E_{slope}$. For both PZ and PZW, the difference between 
$E_{th}$ and $E_{slope}$ is large ($\approx$ 20\%). However, both  potentials shows similar trend i.e decreasing $E$ with increasing 
Al content. 

We have also calculated the elastic constants by using the energy method. In this method the elastic constants are calculated by evaluating the curvature of strain {\em vs} potential
curve~\cite{Yun-Jiang2009,PhysRevB1993,WU2008}. The obtained values match well with the values obtained by explicit deformation method explained in this section.

In summary, the lattice parameters and elastic constants obtained using PZ and PZW potential show similar overall trend;
however, in the case of PZ potential, the Vegard's coefficient is larger as compared to PZW potential.  
In all cases except for $C_{44}$, the trends obtained from the simulations using these two potentials are also the same 
as observed in experiments. Thus, either of these potentials can be used for our simulations. We have carried out
simulations using both these potentials; however, in the main part of the paper we describe the results obtained
using PZW potential. The results are qualitatively the same using PZ potential and are shown in Supplementary Information. 

\subsubsection{System size effects}

\begin{figure}[htpb]
\centering
\includegraphics[trim=0.5cm 0.5cm 0.5cm 0.5cm,height=3in,width=3in]{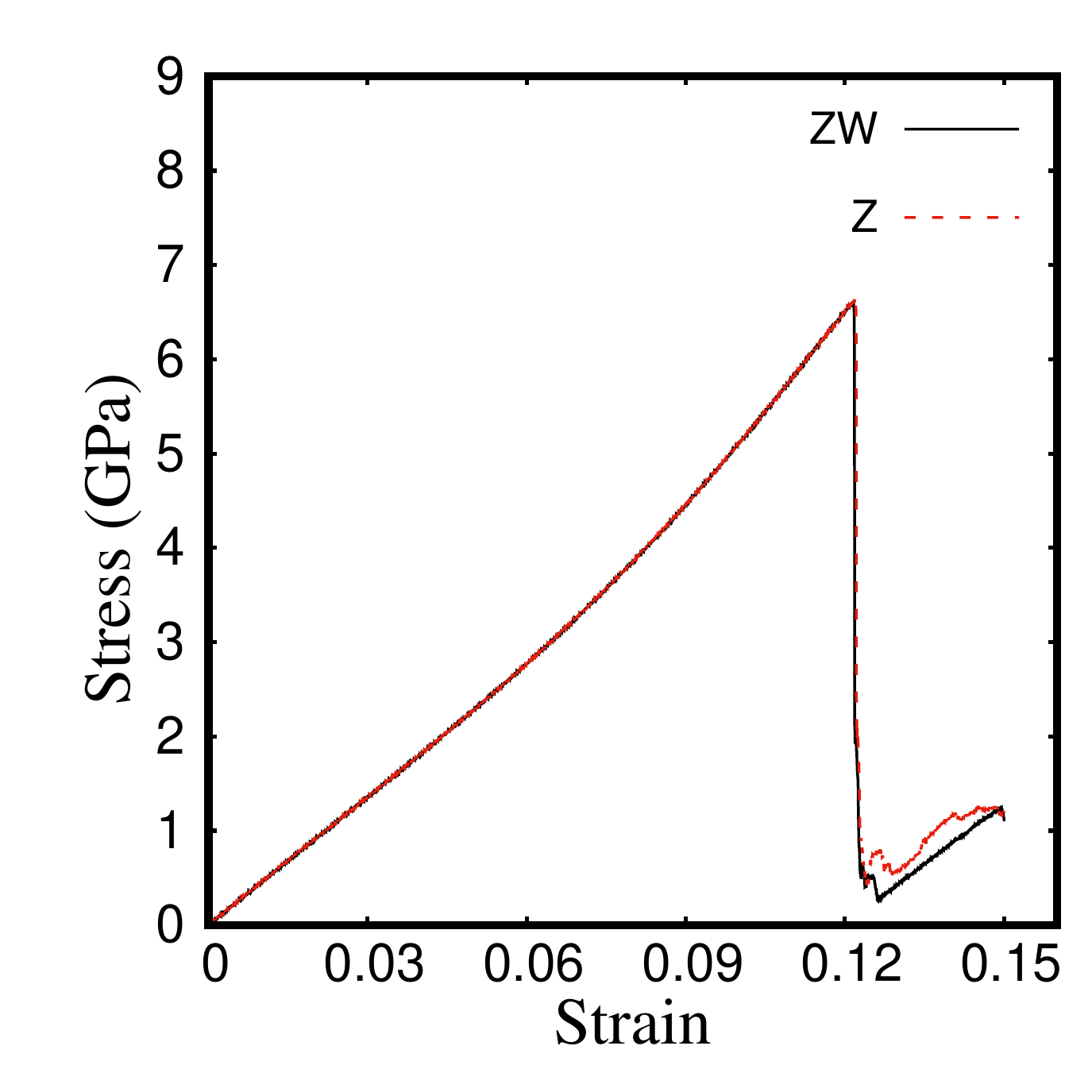} 
\caption{Stress strain response of Cu obtained using Z and ZW potentials for the system size $30\times30\times30$ unit cells.}\label{tensile30u}
\end{figure}

We have carried out tensile simulations using both PZ and PZW potentials (for pure Cu), with the  simulation cells of sizes 
$20\times20\times20$ (32000 atoms) and $30\times30\times30$ (108000 atoms)  unit cells along x, y and z axis respectively. 
In the case of PZ potential, for smaller system size, we have observed that the stress drops to negative values
when dislocation activity starts; in the larger system, this is not observed. On the other hand, PZW potential does not show
any such unphysical behaiour. Hence, we have decided to use PZW for our studies reported in this paper. Having said that,
as shown in Figure.~\ref{tensile30u}, in the case of $30\times30\times30$ unit cells, both the potentials show very 
similar stress-strain behaviour. Hence, we believe that our results will not change even if PZ potential is used as long
as system size effects are avoided. 

In the case of PZW potential, we have also carried out simulations on a system with $40\times40\times40\times$ unit
cells (a system consisting of 256000 atoms). We have found that the stres-strain behaviour is broadly the same. For example,
the dislocation nucleation stress changes at the most by less than 2\% as compared to the system with $30\times30\times30\times$ 
unit cells. Hence, given the time required for the larger simulation and the very small error in results, for the rest of this paper, 
we report results obtained from the system with $30\times30\times30\times$ unit cells.

\section{Evaluation of GSFE curve and SFE} \label{Section3}

Our aim, in this paper, is to correlate the deformation behaviour with SFE. Hence, evaluating the SFE using the
same potentials as used for deformation simulations is essential to rationalise the deformation simulation results
unambiguously. Specifically, in Section~\ref{Section5}, we use a continuum model to evaluate the stress required for 
homogeneous dislocation nucleation. The SFE is one of the inputs to this model along with elastic
moduli, lattice parameters and the size of the nucleated dislocation loop. Further, in the case of Cu-Al
alloys, the nucleation is not always homogenoeus. So, the yield stress observed in MD simulations have
to be explained by also incorporating the decrease in nucleation barrier for heterogeneous nucleation of
partial loops; it is known that this barrier depends on the ratio of unstable and stable stacking fault energies. Hence, in order
to obtain these quantities, we evaluate the GSFE curve (at 0 K) and the SFE. As we describe below, we use the Frenkel-Ladd method
of thermodynamic integration to calculate the SFE as a function of composition and temperature.

\subsection{Calculation of GSFE curve}

The GSFE curve can be generated by displacing one half of the crystal with respect to the other half of the crystal 
and computing the change in potential energy during the displacement~\cite{Ezaz2011,Lu2015,Zimmerman2000,Chassagne2011}. 
Using GSFE curve we can understand the deformation behaviour (i.e slip activity, twin formation and defect interactions) 
and correlate it to the stacking fault energy~\cite{Ezaz2011,Swygenhoven2004}. Slip activity and twin formation can be 
explained in terms of dislocation nucleation and energy barrier associated with it.  Dislocation nucleation can be 
explained using the ratio $\frac{\gamma_{sf}}{\gamma_{usf}}$, where  $\gamma_{sf}$ is stacking fault energy and and $
\gamma_{usf}$ is unstable stacking fault energy. Experimentally only $\gamma_{sf}$ can be measured; however, MD simulations
allow us to access $\gamma_{usf}$ value.

For the calculation of GSFE curve we have used the method described in~\cite{Ezaz2011,Lu2015,Zimmerman2000,Chassagne2011}. 
In these papers, periodic boundary conditions are imposed along x and y axes and along z axis a free surface is considered
(that is ``pps'' boundary conditions).  On the other hand, we have used periodic boundary conditions along all three 
directions (ppp).

\begin{figure}[htbp]
\begin{center}
\subfigure[]{
\resizebox*{3in}{!}{\includegraphics{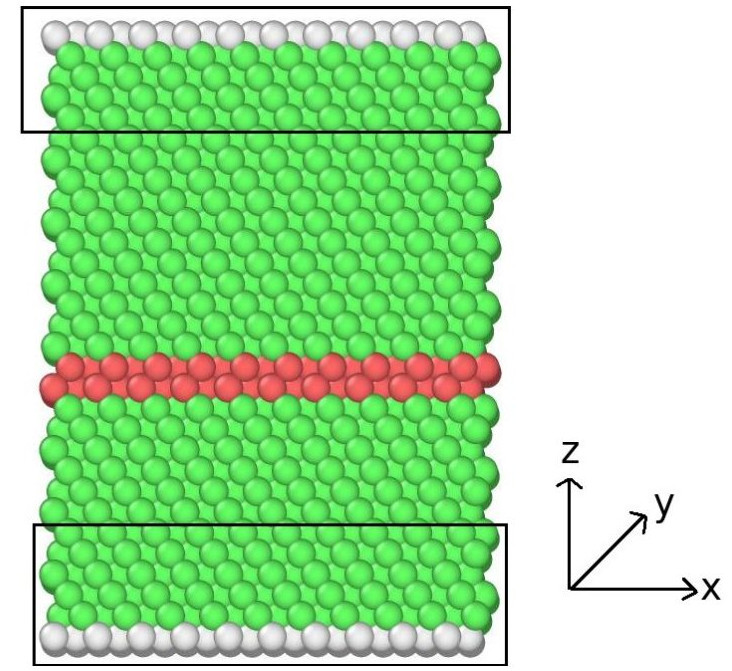}}}\hspace{5pt}
\subfigure[]{
\resizebox*{3in}{!}{\includegraphics{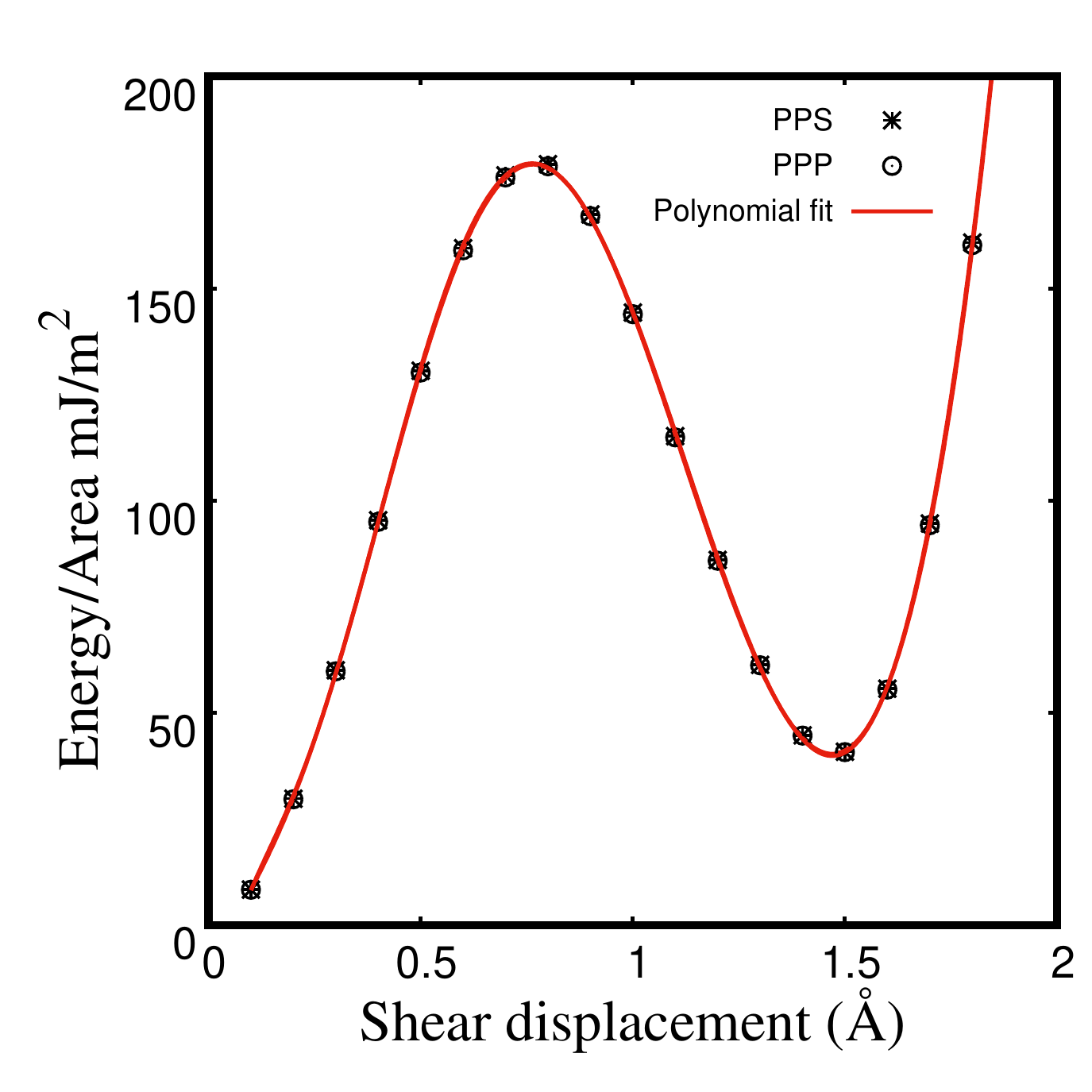}}}
\end{center}
\caption{(a) Microstructure after the displacement of $\frac{a}{\sqrt(6)}$, the top and bottom layers shown in the rectangular box are excluded from the potential enegy calculation (b) GSFE curve for Cu obatained using PPP (periodic in all three directions) and PPS (free surface along z axis) boundary conditions.}\label{BCboth}
\end{figure}

The calculation is done as follows: the x, y and z axes of the simulation box are oriented along $[112]$, $[\bar{1}10]$ and $[\bar{1}\
\bar{1}1]$ crystallographic directions respectively. The simulation box size is $30\times30\times60$ along x, y and z axes. Simulation 
box is divided into two regions i.e top and bottom along z axis. Top region is displaced with respect to bottom region along $[112]$ 
direction in steps of 0.1 \AA  and the potential energy is computed after each displacement. Stacking fault forms at the displacement 
equal to the  magnitude of Shockley partial dislocation ($\frac{a}{\sqrt(6)}$, where $a$ is the lattice parameter). Since periodic 
boundary conditions are used for the calculation, the shear displacement leads to the formation of additional stacking fault at the 
top and bottom, which in turn leaves the system in high energy state. Hence to avoid this problem, 5 atomic layers at the top 
and bottom are excluded while computing potential energy of the system as shown in Fig.~\ref{BCboth}a. 
 
\begin{figure}[htpb]
\centering
\includegraphics[trim=0.5cm 0.5cm 0.5cm 0.5cm,height=3in,width=3in]{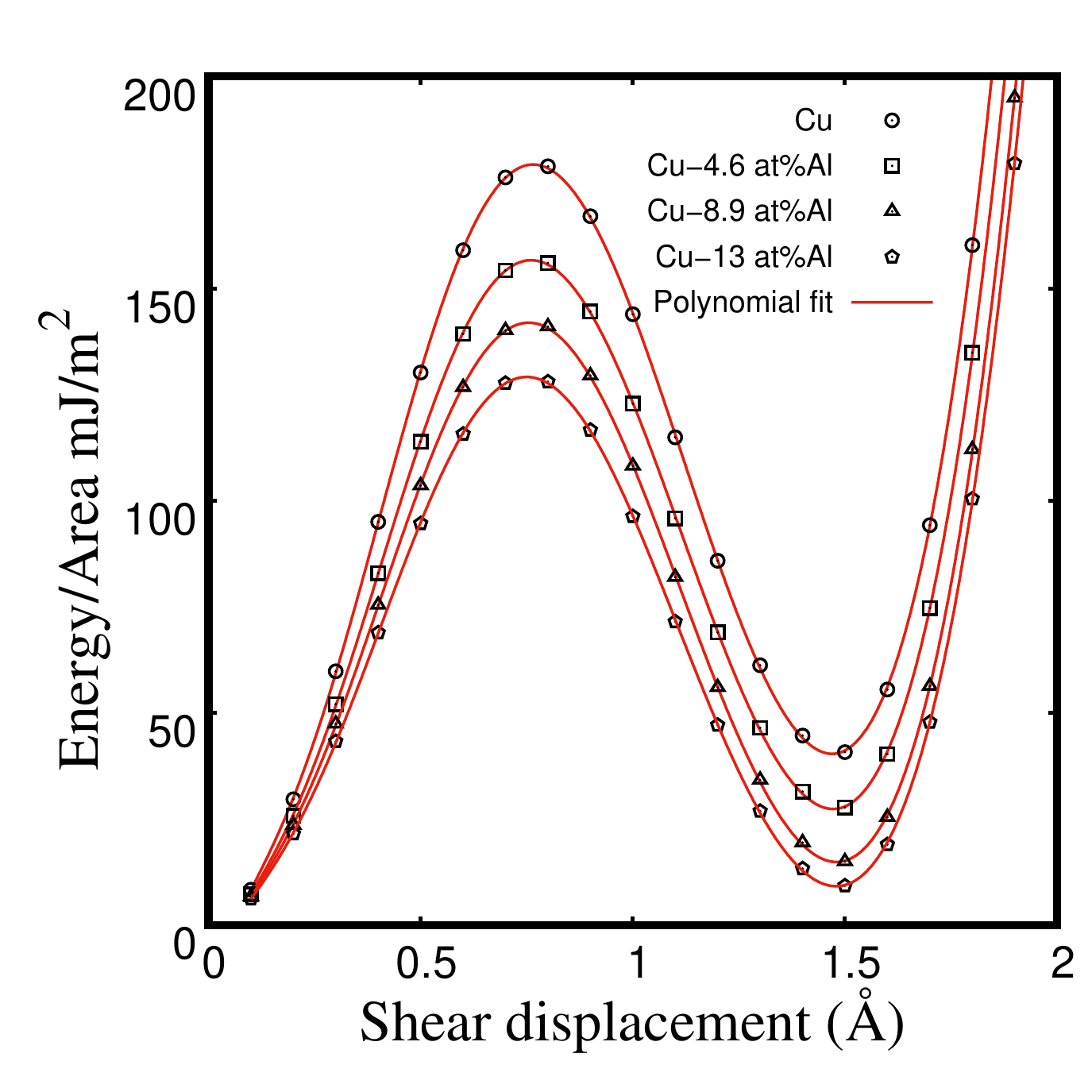} 
\caption{GSFE curve for Cu and Cu-Al alloys.}\label{GSFE}
\end{figure}

We have also calculated the GSFE curve using pps boundary conditions. GSFE curve for Cu obtained using both ppp and pps boundary 
conditions are shown in Fig.~\ref{BCboth}b, both the values are in good agreement. This implies that excluding the 5 atomic layers from 
the potential energy calculation does not affect the values of $\gamma_{sf}$ and $\gamma_{usf}$. Similar procedure is followed to obtain 
GSFE curve for Cu and Cu-Al alloys and is shown in Fig.~\ref{GSFE}, the curves are fitted to $9^{th}$ order polynomial. The values of $
\gamma_{sf}$ and $\gamma_{usf}$ are given in Table~\ref{SFE}. We can see from the table that both unstable stacking fault energy and 
stacking fault energy decreases with increasing Al. The ratio $\frac{\gamma_{sf}}{\gamma_{usf}}$ for pure Cu is 0.23 and it 
matches with the value reported in~\cite{Ezaz2011} (i.e 0.26). The ratio $\frac{\gamma_{sf}}{\gamma_{usf}}$ decreases with increasing Al 
content.

\begin{table}[htbp]
\caption{Unstable stacking fault energy ($\gamma_{usf}$) and  intrinsic stacking fault energy ($\gamma_{sf}$) values (in $\frac{mJ}{m^2}$) of Cu and Cu-Al alloys obtained from MD simulations along with their ratio (which is known to play a key role in heterogeneous nucleation
of dislocations at the or in the vicinity of solute atom sites).}
{\begin{tabular}[l]{@{}lcccccc}\hline

  \footnotesize Material & \footnotesize  {$\gamma_{sf} $ } &  {\footnotesize  $\gamma_{usf} $} & \footnotesize $\frac{\gamma_{sf}}{\gamma_{usf}}$ \\  \hline

  \footnotesize Pure Cu   & \footnotesize   {40.36 $\pm0.0$ }   &\footnotesize   {178.64 $\pm0.0$}&  \footnotesize 0.23 \\                
  \footnotesize Cu-4.6 at.\%Al  &\footnotesize  {27.46 $\pm5.21$ }    &\footnotesize   {158.25 $\pm5.49$} &\footnotesize 0.17 \\                                                         
  \footnotesize Cu-8.9 at.\%Al & \footnotesize   {17.93$\pm3.46$}     & \footnotesize  {143.79 $\pm2.40$}&\footnotesize 0.13 \\                                                         
 \footnotesize  Cu-13.0 at.\%Al & \footnotesize  {12.17 $\pm2.62$}   &\footnotesize  {132.72$\pm3.58$ } &  \footnotesize 0.09\\ \hline
  
 \end{tabular}}
 \label{SFE}
\end{table}

\subsection{SFE in Cu and Cu-Al alloys using thermodynamic integration}

We evaluate the free energy -- using a thermodynamic integration methodology -- of systems with and 
without the stacking faults and the difference in the free energy gives us the stacking fault energy. 
Even though free energy can not be measured directly in atomistic simulations, its derivatives can be
measured and the integration of the derivative gives 
the free energy~\cite{FrenkelSmit}. For example, 

\begin{equation}\label{d1}
 \left( \frac{\partial F }{\partial V}\right) _{N,T} = - P 
\end{equation}

\begin{equation} \label{d2}
 \left( \frac{\partial{F/T}}{\partial{(1/T)}}\right) _{V,N} = E 
\end{equation}

where, F is the free energy of the system with number of particles N and volume V and T, P and E are the temperature, pressure and  
energy   of the system respectively. Pressure and energy can be measured in a simulation. The reversible integration of the above equations, 
from the state under consideration to the state of a known free energy gives the free energy of the system. This methodology is called thermodynamic integration.

In order to calculate the free energy, we also need the free energy of a reference state; in our case, we use the Einstein crystal (that
is, a solid which consists of non-interacting particles coupled to the lattice sites via harmonic springs) as the reference state and 
construct a reversible path using the Einstein crystal. From the Einstein crystal, the real solid is obtained by slowly switching off 
the harmonic springs. Since Einstein crystal has the same crystal structure as that of the solid under consideration, the integration path 
is free of phase transformation and hence reversible.

There are two kinds of approaches to the free energy calculations, namely,  equilibrium and non-equilibrium free energy calculations.
Equilibrium free energy calculation involves thermodynamic integration between  two equilibrium states. The integration path is reversible 
and passes through quasi-static processes and is time independent. In non-equilibrium free energy calculation, the integration path passes 
through time dependent processes. In order to understand this, let us consider an equation, which gives the free energy difference between two 
thermodynamic states~\cite{Freitas2016}.
 
 \begin{equation}\label{deltaF}
  \Delta F = F(\lambda_f)-F(\lambda_i) = \int_{\lambda_i}^{\lambda_f} \langle{\frac{\partial H}{\partial{\lambda}}}\rangle_{\lambda} d \lambda
 \end{equation}
 
where   $F(\lambda_i)$ and $F(\lambda_f)$ are free energies  of initial and final states respectively, $\lambda$ is switching parameter and  H is Hamiltonian of the system. In equilibrium approach, the integral in the above equation  is evaluated  in terms of reversible work done ($w^{rev}_{i-f}$) along the path between  thermodynamic states. But, in non-equilibrium approach the integral is evaluated  in terms of irreversible work done ($w^{irrev}_{i-f}$) along the path and $ \lambda = \lambda(t)$. Since non-equlibrium approach is efficient compared to equilibrium approach we use non-equilibrium thermodynamic integration for free energy calculation. We use Frenkel-Ladd (FL) path to calculate free energy at 300 K at which the deformation simulations are carried out.

\subsubsection{Calculation of free energy using FL path}

The methodology of FL path for free energy evaluation is described in detail in~\cite{Freitas2016} and references
therein. In this subsection, we give a brief description for the sake of completeness.
 
 The Hamiltonian given in Eq.~\ref{deltaF} as a function of switching parameter $\lambda$ can be written as:

 \begin{equation}\label{Hamiltonian}
 H(\lambda) = \lambda H_f + (1-\lambda)H_i
 \end{equation}  
 
 where $H_f$ and $ H_i$ represents the Hamiltonian of final and initial states respectively. Here we consider $H_i$ as the Hamiltonian of the real solid corresponding to switching parameter $\lambda$ = 0; 
 and  $H_f$ as the Hamiltonian of the reference state (that is,  Einstein crystal) corresponding to switching parameter $\lambda$ = 1. $H_f$ and $ H_i$ are given as follows:

 \begin{equation}\label{Hreal}
 H_i = \Sigma_{n=0}^{N}\frac{\mathbf P_{i}^2}{2m} + U(\mathbf r)
 \end{equation}
 
 \begin{equation}\label{HEinstein}
 H_f = \Sigma_{n=0}^{N}\frac{\mathbf P_{i}^2}{2m} + \frac{1}{2}m\omega^2(\mathbf r_i - \mathbf r_{i}^0)^2
 \end{equation}

where, N is the number of atoms, m is the mass of each atom, $U(\mathbf r)$ is the potential energy of the system, $\omega$ is the oscillation frequency of the Einstein crystal and  $ \mathbf r_{i}^0$ is the equilibrium position of $i^{th}$ atom. Now the free energy difference between the two systems  given in the Eq.~\ref{deltaF} for the switching from $\lambda$ = 0 to 1, can be written as:

 \begin{equation}\label{deltaF1}
  \Delta F = F(\lambda_f)-F(\lambda_i) = \int_{0}^{1} d \lambda \langle{H_f - H_i}\rangle_{\lambda} 
 \end{equation}

 The irreversible work done during the switching in switching time $t_s$ is given as:
 
  \begin{equation}\label{W1}
 W_{0-1}^{irr} = \int_{0}^{t_s} dt \frac{d\lambda}{dt} \left[ H_f(\Gamma(t))-H_i(\Gamma(t))\right] 
 \end{equation}

where $\Gamma(t)$ represents the phase space trajectory of the system during switching. Free energy of the real solid (system of our interest) is obtained by computing the irreversible work done during switching from $\lambda$ = 0 to 1 ($W_{0-1}^{irr}$ -- forward integration) and the irreversible work done during switching from $\lambda$ = 1 to 0 ($W_{1-0}^{irr}$ -- backward integration) by using the following equation:

\begin{equation}\label{FE}
 F_0(N,V,T) = F_E(N,V,T) + \frac{1}{2}(\overline{W_{i-f}^{irr}} - \overline{W_{f-i}^{irr}})
\end{equation}
where, $F_E(N,V,T)$ is free energy of the Einstein crystal :
$F_E(N,V,T) = 3N k_BT\ln \left( \frac{\hslash \omega}{k_B T} \right) $, N is total no of atoms, $k_B$ is Boltzmann constant, $T$ is  temperature,  $ \hslash   $ is Planck's constant and $\omega$ is oscillator frequency ($\omega$ = $\sqrt(\frac{K}{m})$, $K$ is spring constant and $m$ is mass of an atom). 

Methodology to calculate free energy of the system using FL path  is as follows:

\begin{enumerate}

\item Calculate the lattice parameter at temperature of interest.

\item Compute mean square displacement $(\left\langle (\Delta r)^2\right\rangle )$ and hence the spring constant. 

\item Carry out thermodynamic integration using \emph{fix ti/spring} command of LAMMPS.

\item Repeat the integration several times to get average $W^{irr}$ i.e $\overline{W_{i-f}^{irr}}$ and $\overline{W_{f-i}^{irr}}$.

\item Calculate the free energy by using the Eq.~\ref{FE}

\end{enumerate}

In Table~\ref{SFE1} we summarise the results of our simulations. In 
Supplementary Information, we have given the scripts used
by us.

\begin{table}[htbp]
\caption{$\gamma_{sf}$ of Cu and Cu-Al alloys at 300 K calculated using thermodynamic integration. The 
experimentally determined SFE values are also given for an easy comparison.}
{\begin{tabular}[l]{@{}lcccccc}\hline

  \footnotesize {Material} & \footnotesize {$\gamma_{sf}$ in mJ/m$^2$ } & Experimental \\
  && literature (\footnotesize {$\gamma_{sf}$ in mJ/m$^2$ })  \\  \hline

  \footnotesize Pure Cu   & \footnotesize  {23.59 $\pm0.0$ }   &  \footnotesize{43~\cite{CarterRay}} \\                
  \footnotesize Cu-4.6 at.\%Al  &\footnotesize {20.72 $\pm1.18$ }   & \footnotesize{25~\cite{Rohatgi}} \\                                                         
  \footnotesize Cu-8.9 at.\%Al & \footnotesize  {17.87$\pm0.59$}   &\footnotesize{13~\cite{Rohatgi}}  \\                                                         
 \footnotesize  Cu-13.0 at.\%Al & \footnotesize {10.46 $\pm0.91$}  &  \footnotesize {6~\cite{Rohatgi}} \\ \hline
  
 \end{tabular}}
 \label{SFE1}
\end{table} 

\section{Tensile deformation behaviour of Cu and Cu-Al alloys} \label{Section4}

\begin{figure}[htbp]
\begin{center}
\subfigure[]{
\resizebox*{3in}{!}{\includegraphics{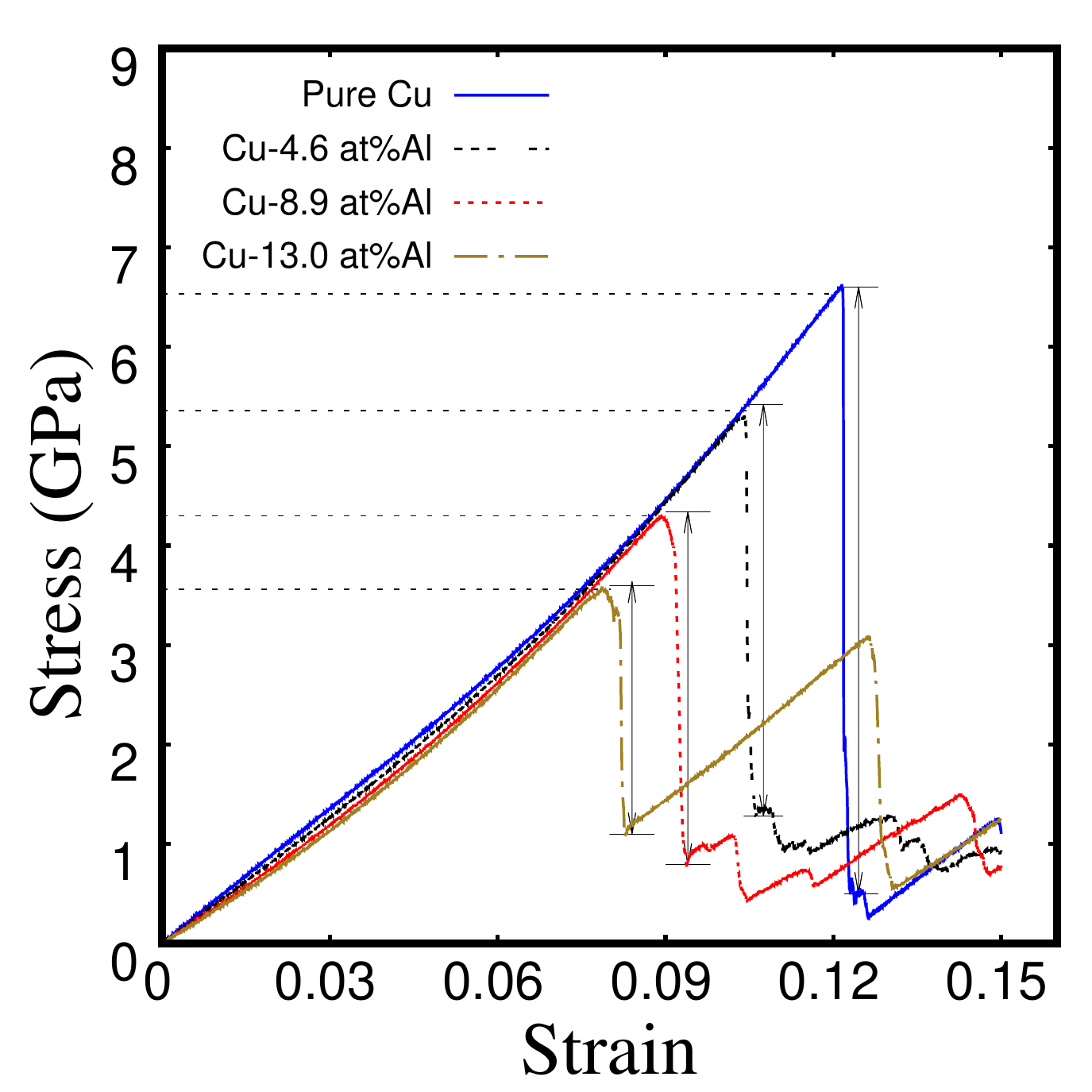}}}\hspace{5pt}
\subfigure[]{
\resizebox*{3in}{!}{\includegraphics{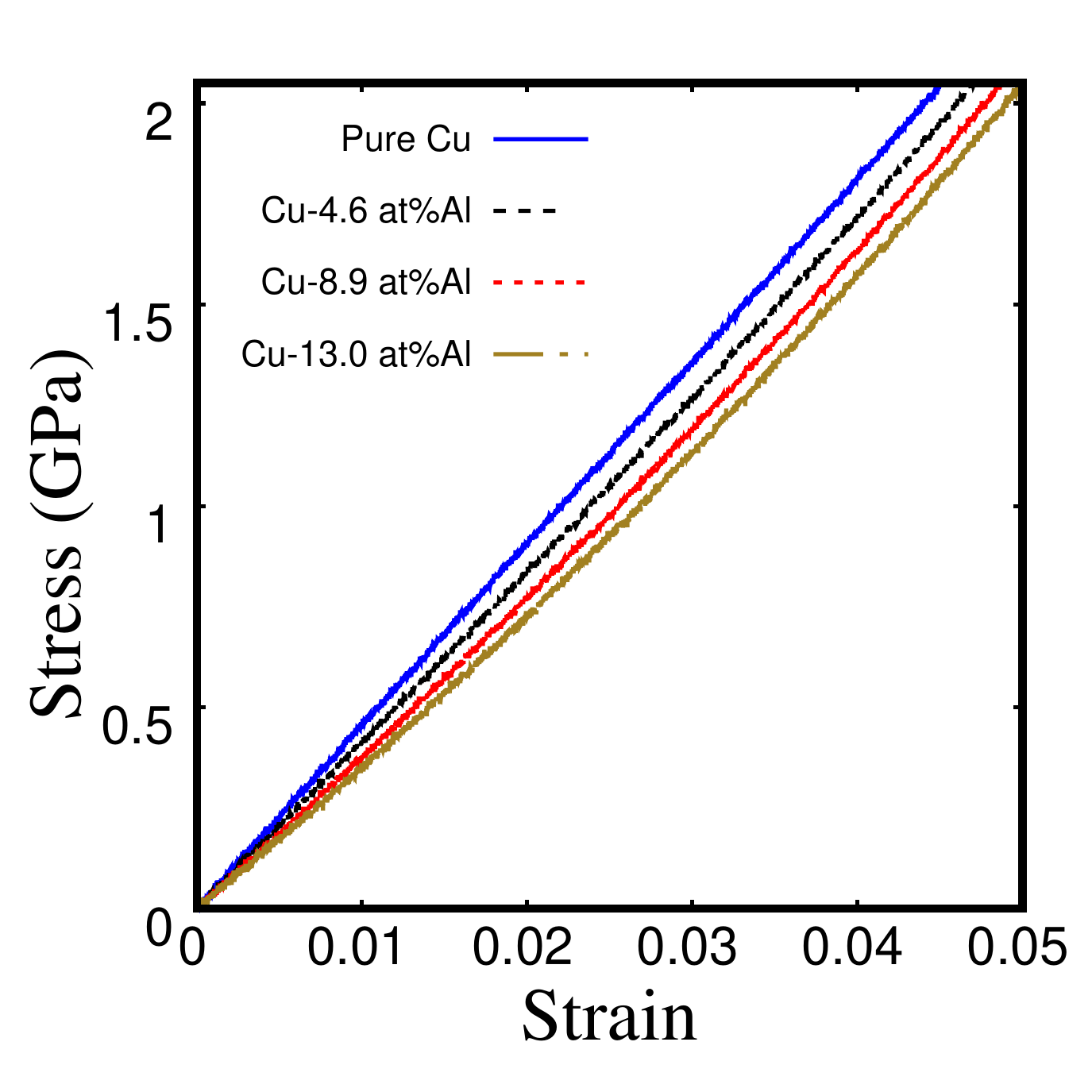}}}
\end{center}
\caption{(a) Stress-strain response of Cu and Cu-Al alloys, black dotted lines show the yield point and black arrows show the amount of drop in stress at yield. (b) Initial portion of the stress strain plot.}\label{stress-strainZWard}
\end{figure}

The representative stress strain response for pure Cu, Cu-4.6 at.\%Al, Cu-8.9 at.\%Al and Cu-13.0 at.\%Al alloys is shown in Fig.~\ref{stress-strainZWard}a. From figure we can see that following the initial portion there is an abrupt drop in the stress. In the figure, in all plots, we have marked the first abrupt drop in stress by black dotted lines. The stress at these points decreases with increasing Al content. Further, the magnitude of drop (shown by black arrows) also decreases with increasing Al content. 

The initial portion of the stress strain curves for Cu and Cu-Al alloys is shown in Fig.~\ref{stress-strainZWard}b. From figure we can see that the slope of the stress-strain curve (Young's modulus) varies with Al composition.

In this section we discuss in detail about the variation of Young's modulus, drop in stress and the magnitude of drop in stress with Al composition.


\begin{table} 
\caption{Variation of yield stress ($\sigma_y$), drop in stress ($\Delta \sigma$) and Young's modulus ($E_{slope}$) for Cu and Cu-Al alloys, in units of GPa}
 \begin{tabular}[l]{@{}lcccccc} \hline
 \footnotesize Material & \footnotesize $\sigma_y$  &   \footnotesize $\Delta{\sigma}$  & \footnotesize $E_{slope}$   \\ \hline

 \footnotesize Pure Cu           & \footnotesize  6.58$\pm0.02$       & \footnotesize  6.08$\pm0.02$ & \footnotesize 45.37$\pm 0.16 $   \\
 \footnotesize Cu-4.6 at.$\%$Al & \footnotesize  5.29$\pm0.04$       & \footnotesize  4.22$\pm0.18$  & \footnotesize 41.72$\pm  0.24$   \\
 \footnotesize Cu-8.9 at.$\%$Al & \footnotesize  4.13$\pm0.16$       & \footnotesize  3.29$\pm0.20$   &\footnotesize 38.55$\pm  0.30$  \\
 \footnotesize Cu-13.0 at.$\%$Al& \footnotesize  3.52$\pm0.05$       & \footnotesize  2.45$\pm0.60$   & \footnotesize  36.17$\pm 1.59 $   \\ \hline

 \end{tabular}
 \label{YSandDropTable}
\end{table}

\subsection{Variation of Young's modulus}

We have calculated the Young's modulus as the slope of the initial portion of the stress-strain plot by linear fitting till 2.\% strain. The values of Young's modulus for Cu and Cu-Al alloys are given in Table~\ref{YSandDropTable}. From table we can see that the Young's modulus decreases with increasing Al content; which implies that the addition of Al to Cu makes the material more compliant. Experimental measurements on polycrystalline Cu-Al alloys~\cite{Lenkkeri1973, Koster} does indeed show that the 
Young's modulus decreases with alloying and hence these results are consistent with the experimental
observations.

MD simulations performed on other binary alloys show that the elastic stiffness either decreases or increases depending on the solute added~\cite{TJRupert}. Rupert observed that the Young's modulus of Cu decreases with the addition of more compliant solute and increases with the addition of stiffer solute~\cite{TJRupert}. Thus, adding Pb and Sb~\cite{RKRajgarhia} to Cu leads to
a reduction in the Young's modulus of the alloys compared to copper. In our case also, the addition 
of Al which is more compliant than Cu reduces the Young's modulus.  

\subsection{Variation of yield stress}

\begin{figure}
\begin{center}
\includegraphics[scale=0.4]{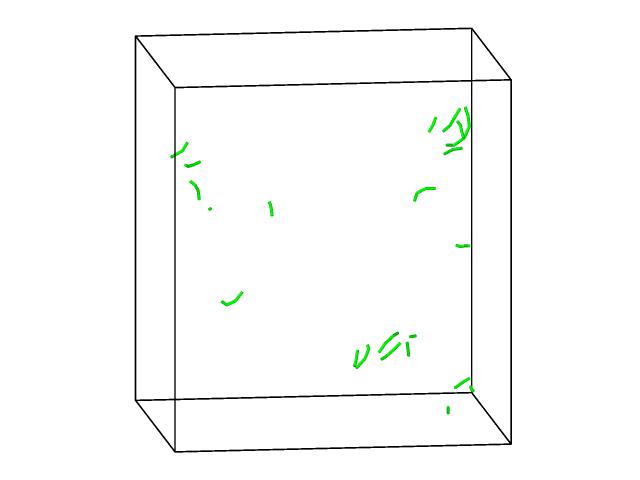}
\includegraphics[scale=0.4]{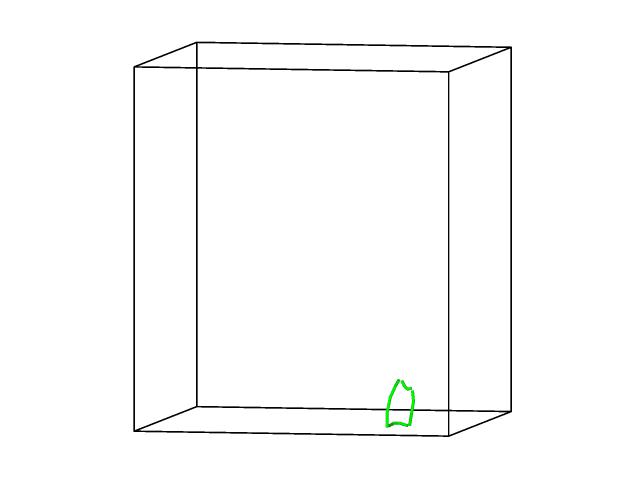}
\includegraphics[scale=0.4]{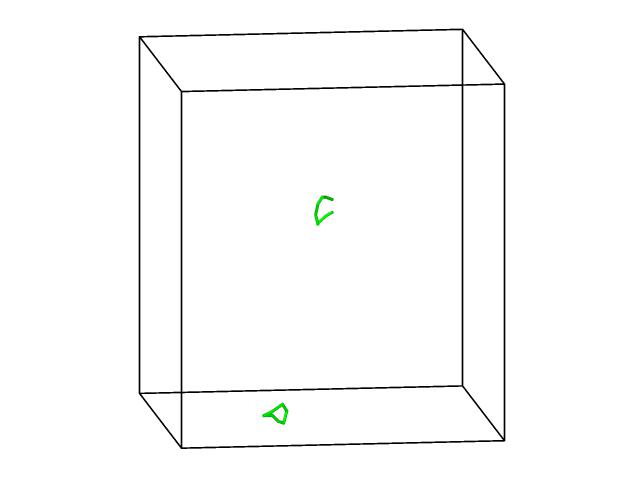}
\includegraphics[scale=0.4]{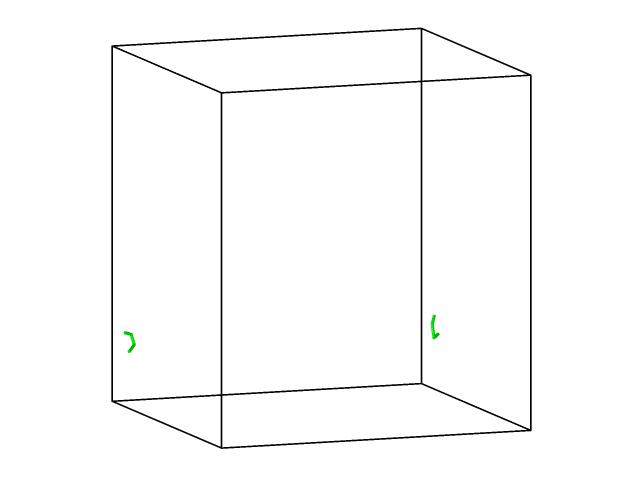}
\caption{Nucleation of Shockley partial loop in (a) pure Cu,  (b) Cu-4.6 at.\% Al, (c) Cu-8.9 at.\%Al and
(d) Cu-13.0 at.\%Al.}\label{LoopFormation}
\end{center}
\end{figure}

From stress strain plot shown in Fig.~\ref{stress-strainZWard}a, we can see that there is an abrupt drop in the stress. The microstructures of pure Cu and Cu-Al alloys at these points are shown in Figures.~\ref{LoopFormation}a-d. In the microstructure  green lines indicate the Shockley partial dislocations. From the microstructures we can see that the yield point corresponds to nucleation of Shockley partial dislocations. The stress required for dislocation nucleation can thus be called 
as yield stress~\cite{Salehinia2014,Zhao2009} (and the corresponding stress as yield stress). 

\begin{figure}[htbp]
\begin{center}
\subfigure[]{
\resizebox*{3in}{!}{\includegraphics{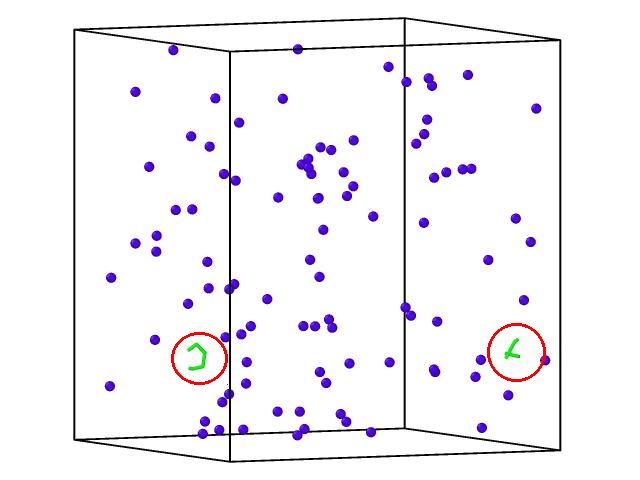}}}\hspace{5pt}
\subfigure[]{
\resizebox*{3in}{!}{\includegraphics{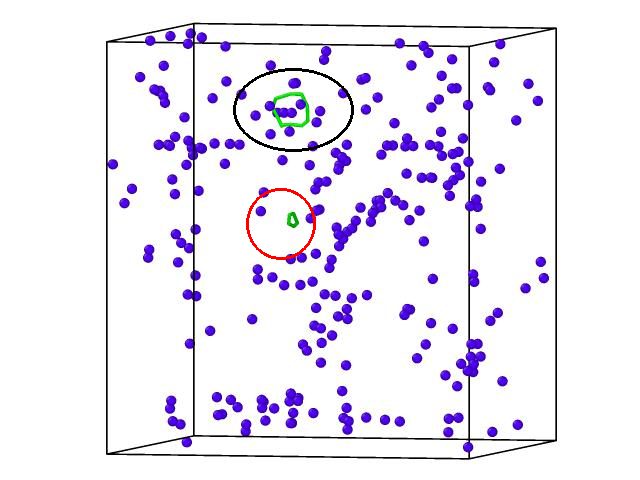}}}
\end{center}
\caption{Nucleation of Shockley partials (shown in green -- using dislocation analysis (DXA) of OVITO) 
in (a) Cu-4.6 at.\% and (b) Cu-8.9 at.\%Al systems. The blue circles are the Al atoms; for the sake of 
clarity, copper atoms are not shown. As is clear from the figure, both homogeneous and heterogeneous 
nucleation are seen. }\label{hetero}
\end{figure}

In pure Cu, dislocation nucleation is always homogeneous.  In Cu-Al, where Al atoms can act as sites for
 dislocation nucleation we have observed both homogeneous and heterogeneous nucleation. Figure.~\ref{hetero} shows nucleation of Shockley partial dislocations in Cu-4.5 at.\%Al and Cu-8.9 at.\%Al system. In the microstructure Al atoms are coloured in blue, homogeneous nucleation is marked by red circle and heterogeneous nucleation near Al atoms is marked by black circle. From the figure we can observe that in Cu-4.5 at.\%Al, nucleation is homogeneous and in Cu-8.9 at.\%Al,  nucleation is both homogeneous and heterogeneous.

In experiments which are typically done on polycrystalline materials where dislocations pre exist before loading, yield stress is defined as the stress required to move the dislocations (which is of the order of few MPa). But in MD simulations, load is applied on defect free single crystals and hence the deformation is dislocation nucleation limited; and the nucleation stress is of the order of few GPa~\cite{Tschopp2008,Salehinia2014,Zhao2009}. 
 
\subsection{Variation of drop in stress}

\begin{figure}[htpb]
\centering
\includegraphics[trim=0.5cm 0.5cm 0.5cm 0.5cm,height=3in,width=3in]{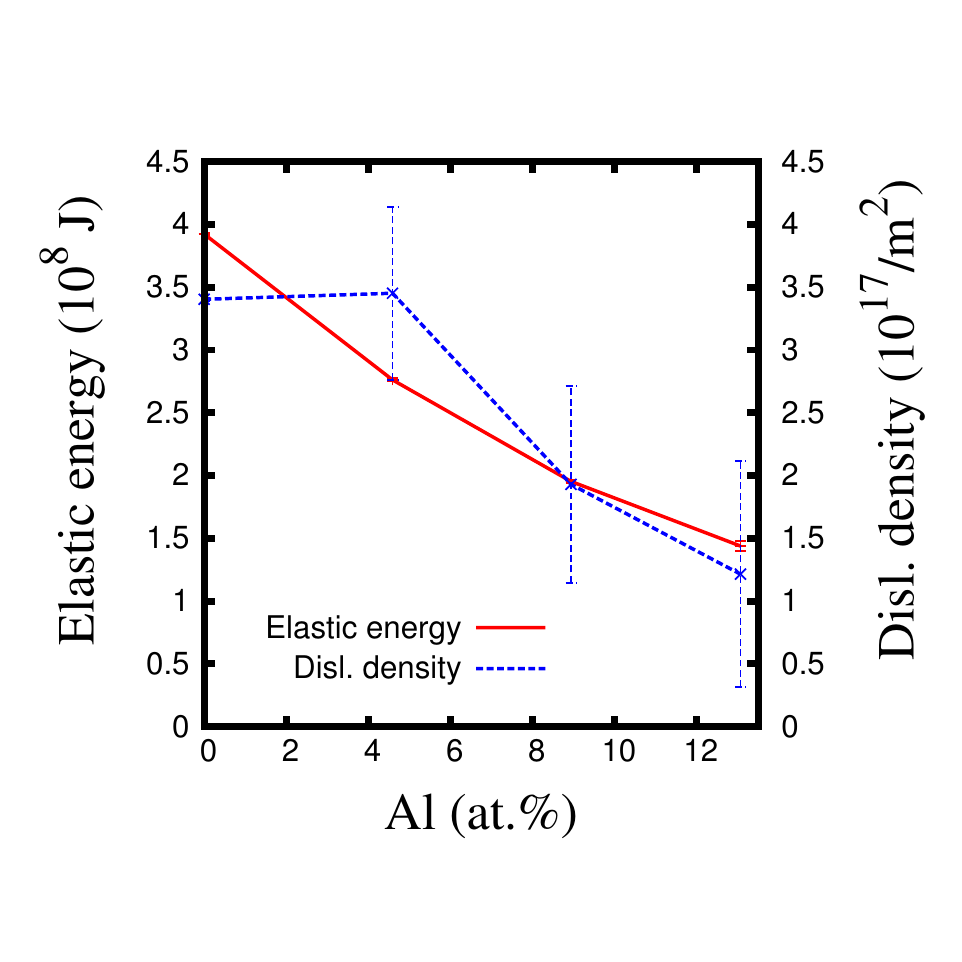} 
\caption{Variaiton of dislcoation density with Al composition after the first yiled drop. }\label{density1}
\end{figure}

As we show in the next section, in addition to ease of heterogeneous nucleation, the homogeneous
nucleation of Shockley partial dislocations becomes easier with Al addition due to decreasing SFE; hence 
the yield stress decreases with increasing Al. It can also be seen from Fig.~\ref{stress-strainZWard}a that the strain at yield also decreases. Further, with the addition of Al, the material becomes more compliant. Therefore, the stored elastic energy decreases with Al addition. We have calculated the stored elastic energy at yield as $\frac{1}{2}E_{slope}\varepsilon^2$ where $\varepsilon$ is the strain at yield and $E_{slope}$ is the Young's modulus calculated from initial portion of the stress strain plots shown in Fig.~\ref{stress-strainZWard}a. The variation of the stored elastic energy with Al composition is shown in Fig.~\ref{density1}. We have also calculated the elastic energy using the stress strain response given in~\cite{Setoodeh2008,Spearot2009,Yang2009,Sainath2017,Farkas2009}, the order of elastic energy calculated from our simulations are in good agreement with those values. It is clear from the figure that the stored elastic energy decreases with increasing Al content; hence at yield point system has less energy to release in the form of dislocations. Thus, the magnitude of drop in stress decreases with Al addition. 

\begin{figure}
\begin{center}
\includegraphics[scale=0.25]{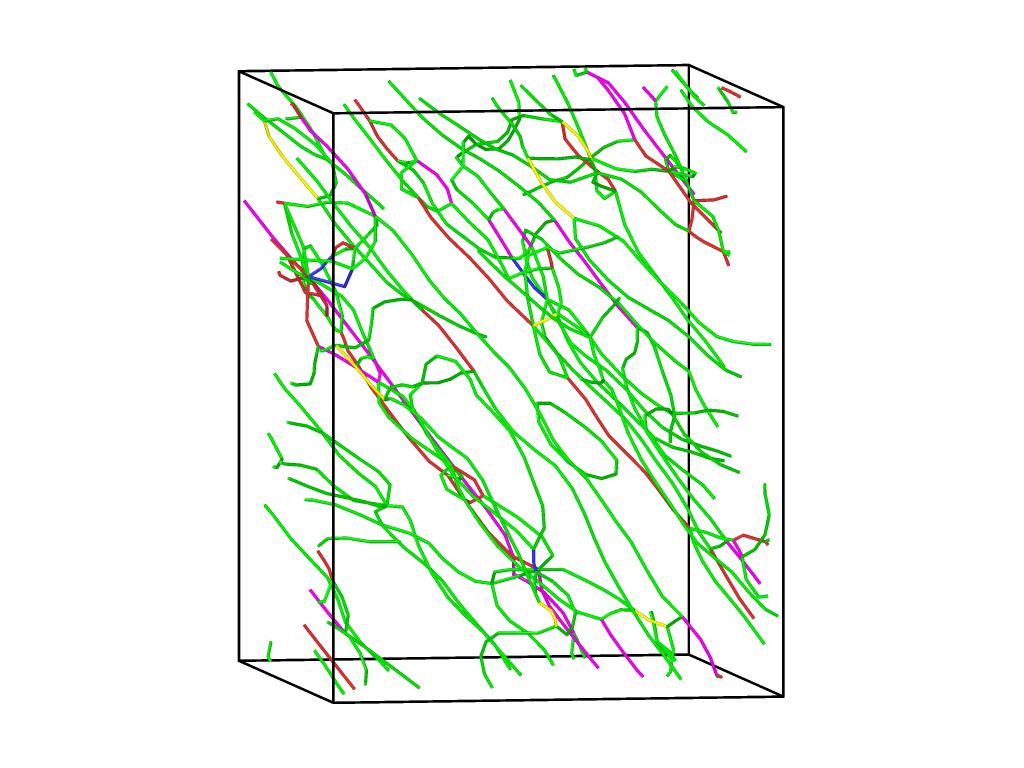}
\includegraphics[scale=0.25]{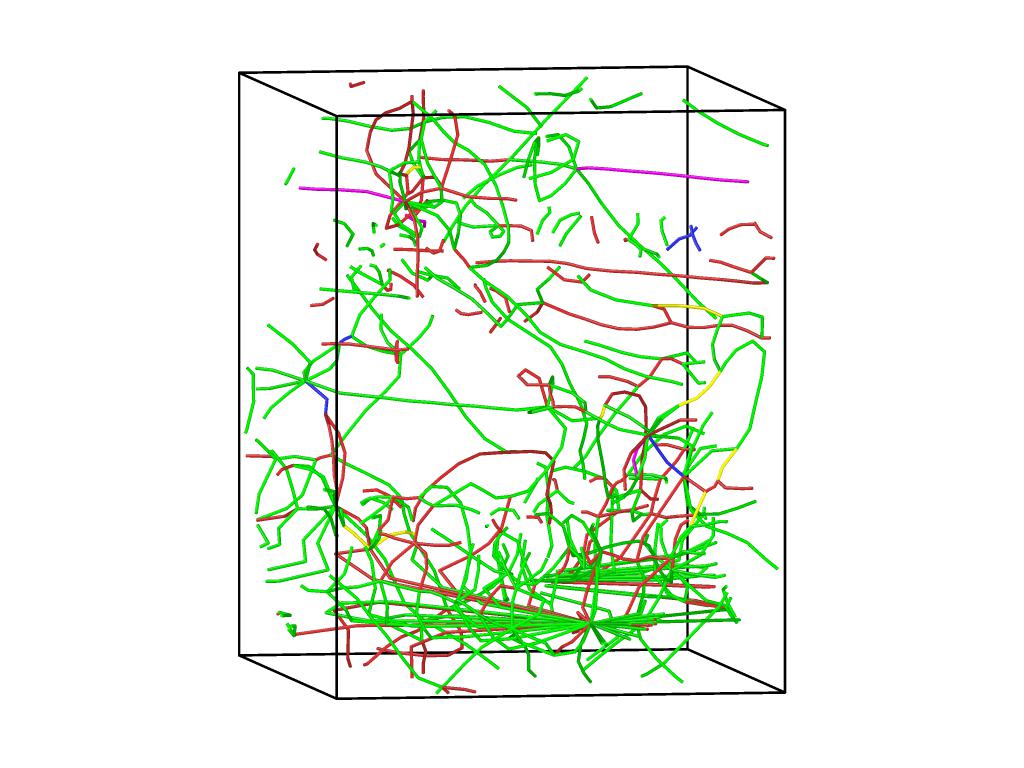}
\includegraphics[scale=0.25]{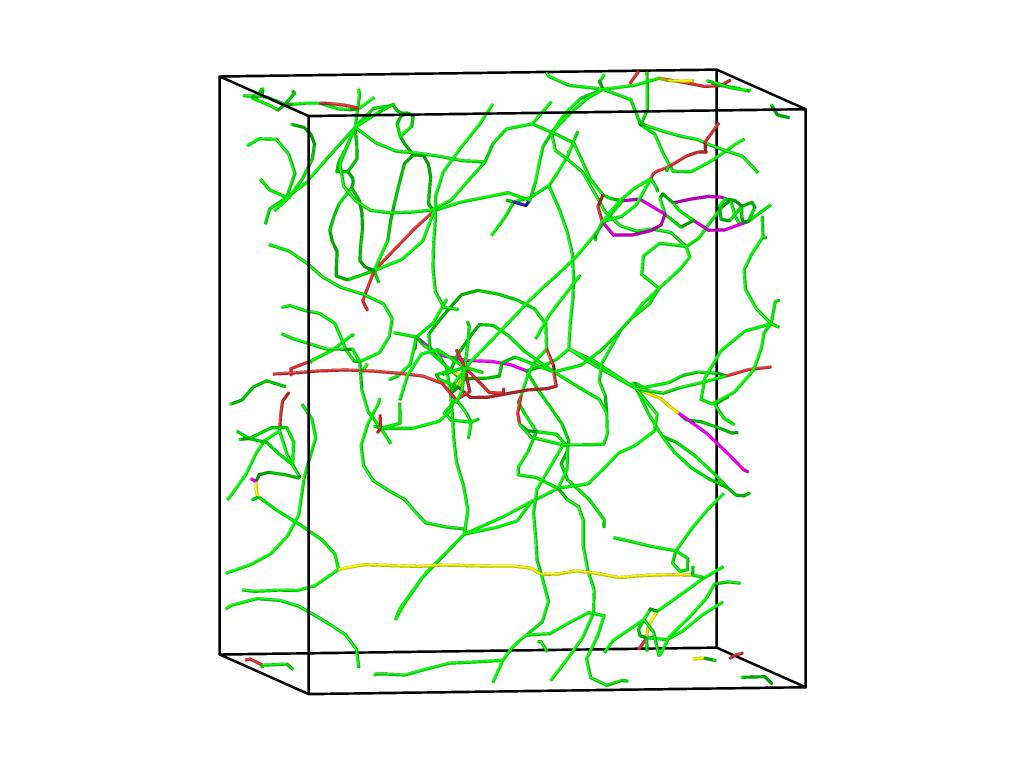}
\includegraphics[scale=0.25]{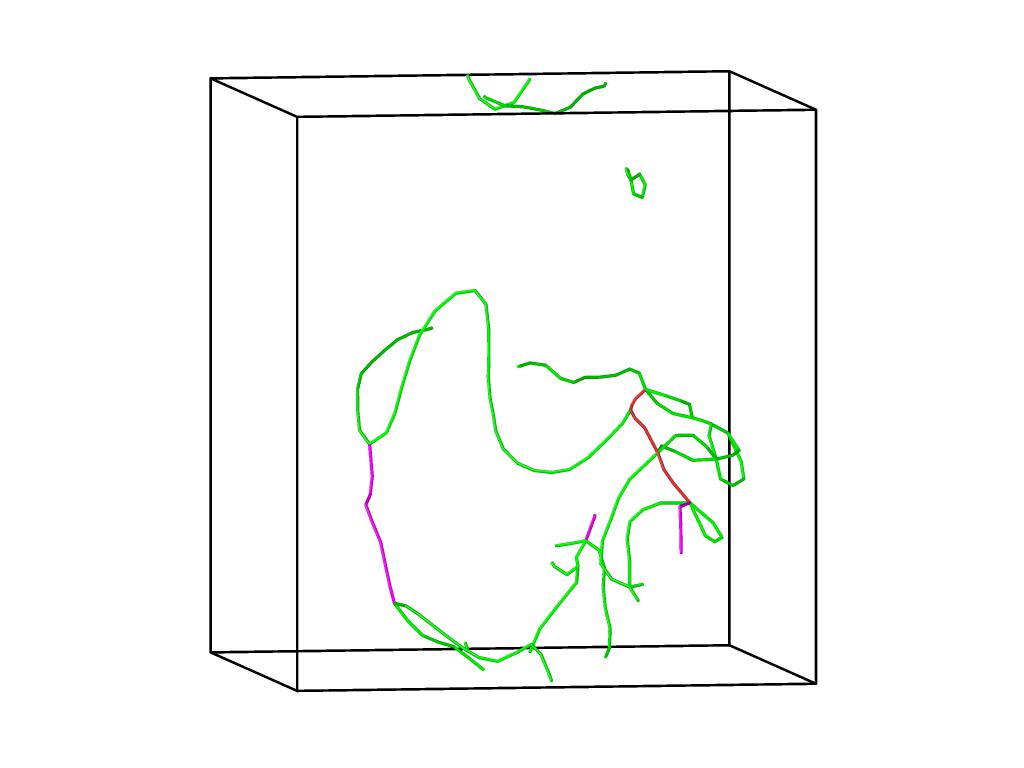}
\caption{Dislocation structure after the first yiled drop in (a) pure Cu,  (b) Cu-4.6 at.\% Al, (c) Cu-8.9 at.\%Al and(d) Cu-13.0 at.\%Al.}\label{density2}
\end{center}
\end{figure}

We have calculated the dislocation density after the first yield drop and shown in Fig.~\ref{density1}. Since the stored elastic energy decreases with Al addition, dislocation density after the first yield drop decreases with increasing Al content. The dislocation structure at those points for Cu-Al alloys are shown in Fig.~\ref{density2}a-d. Here, dislocation density is calculated by dividing total length of the dislocations by volume of the simulation cell. The total length of the dislocations is obtained from the visualisation tool, Ovito. The calculated dislocation density values from our simulations are of the 
order $10^{17}/m^2$. Note that this value is higher than the typical dislocation density of $10^{14}$- $ 10^{15} /m^2$~\cite{An2009,Gubicza2007,Rifai2018,Rohatgi2002,Rohatgi2001a,Tao2013}. The reason for this 
difference, we believe, is the very high strain rates in MD simulations ($10^{8}$ $s^{-1}$); it is known that that the dislocation density increases  with increasing strain rates~\cite{Ye2019}.

The total dislocation density after the first yield point is
related to the stored elastic energy. The change in the stored elastic energy
with Al additions consists of two contributions; one is the change in the
modulus; the second is the change in yield stress (through change in SFE).
Since both contributions lead to a decrease in stored energy, they lead to a decrease in dislocation density.

\section{Discussion} \label{Section5}

We have seen that the onset of plasticity in our simulations occur by the nucleation of Shockley partials.
These nucleations are homogeneous in pure copper; even in Cu-Al alloys, there is some homogeneous
nucleation. Hence, we have used the 
continuum model proposed by Aubry et al.~\cite{Sylvie} to estimate the stress required for the homogeneous nucleation of Shockley partial dislocation loops. In the following section we discuss in detail about the calculation of nucleation stress using the continuum model.

\subsection{Stress required for homogeneous dislocation nucleation}

Following  Aubry et al., consider a partial dislocation loop of radius $R$ and the core radius $r$ in a material with shear modulus $\mu$ and Poisson's ratio $\nu$. The elastic strain energy of partial dislocation loop with Burger's vector $\mathbf{b}$ is given by the equation:

\begin{multline}\label{disl-nucleation}
E_{elastic} =  2 \pi R \frac{\mu b^{2}}{8 \pi} \left\lbrace \frac{2-\nu}{1-\nu}\left[\ln\left({\frac{8 R}{r}}\right)-2\right]
 +0.5\right\rbrace
\end{multline}

For Shockley partial loop which encloses stacking fault of area A, the stacking fault energy term is added to the equation~\ref{disl-nucleation}. 

\begin{multline}\label{disl-nucleation1}
E =  2 \pi R \frac{\mu b^{2}}{8 \pi} \left\lbrace \frac{2-\nu}{1-\nu}\left[\ln\left({\frac{8 R}{r}}\right)-2\right]
 +0.5\right\rbrace + \\  + \gamma_{sf} A 
\end{multline}

Work done by the applied shear stress $\tau$ acting on the glide plane of dislocation loop to nucleate the partial loop of area A with Burger's vector $\mathbf{b}$ is $b \tau A$. The energy barrier for the dislocation nucleation is zero when the work done by the applied shear stress $\tau$ is equal to the sum of the elastic energy of the partial dislocation loop and the stacking fault energy term.  Therefore, the shear stress required for dislocation nucleation $\tau_{nuc}$ is obtained by the equation:

 \begin{multline}\label{FinalEq1}
\tau_{nuc} =    2 \pi R \frac{\mu b}{8 \pi A} \left\lbrace \frac{2-\nu}{1-\nu}\left[\ln\left({\frac{8 R}{r}}\right)-2\right]
 +0.5\right\rbrace + \gamma_{sf} 
\end{multline}
 
By replacing $A$ by $\pi R^2$,  $\mu$  by $C_{44}$ and  $\nu$ by $\frac{C_{12}}{C_{11}+C_{12}}$ in the above equation we get:  

\begin{equation}\label{FinalEq}
  \tau_{nuc} =  \frac{C_{44} b}{4 \pi R} \left\lbrace \frac{2C_{11}+C_{12}}{C_{11}}\left[\ln\left({\frac{8 R}{r}}\right)-2\right]
 +0.5\right\rbrace+\frac{\gamma_{sf}}{b}.
\end{equation}


Using the above equation we can calculate the shear stress required for homogeneous nucleation of dislocations. The corresponding tensile stress $\sigma_{nuc}$  for [001] loading direction can be calculated as {$\sigma_{nuc}$ = $\frac{\tau_{nuc}}{\cos(\phi)\cos(\lambda)}$}. Here, $\phi$ and $\lambda$ are the angles made by the slip plane normal and the Burgers vector, respectively with the loading direction. 

The inputs required for the calculation of nucleation stress are: lattice parameters (for Burger's vector calculation), elastic constants ($C_{11}$, $C_{12}$, $C_{44}$) and fault energies. We have used the values of these inputs obtained from the MD simulations. We have used the dislocation core radius ($r$) and radius of the dislocation loop ($R$) as fitting parameters. The fitting parameters were chosen so as to match the calculated values of $\sigma_{nucl}$ to simulation values of $\sigma_{y}$ of Cu and the same values of $R$ and $r$ are used for the calculation of $\sigma_{nucl}$ of Cu-Al alloys. The fitting parameters used are: $r= b$ and $R= 2$ nm.  The fitted value of $R$ matches the radius of dislocation loop observed in some of our simulations. Also the value of core radius we have used lies in the range b-5b, which is the typical value of core radius~\cite{HullBacon,WandW}. 

Since tensile deformation simulations are performed at 300 K, in the calculation of nucleation stress we need to consider the thermal contribution to the stress in the model.
Aubry et al., report that the continuum model can be used to calculate the
stress required for dislocation nucleation at finite temperatures by incorporating the generalised stacking fault energy curves at the temperature of
interest. Also, Warner et al.~\cite{Warner2009} report that the energy barrier for dislocation nucleation is temperature dependent and it arises from the temperature dependent material parameters i.e lattice parameter, elastic constants and stacking fault energy. Hence in our calculations temperature dependence of nucleation stress  ($\sigma_{nuc}$) at 300 K is incorporated by using the values of elastic constants, lattice parameters and stacking fault energy computed at 300 K from MD simulations.

\begin{figure}[htpb]
\centering
\includegraphics[trim=0.5cm 0.5cm 0.5cm 0.5cm,height=3in,width=3in]{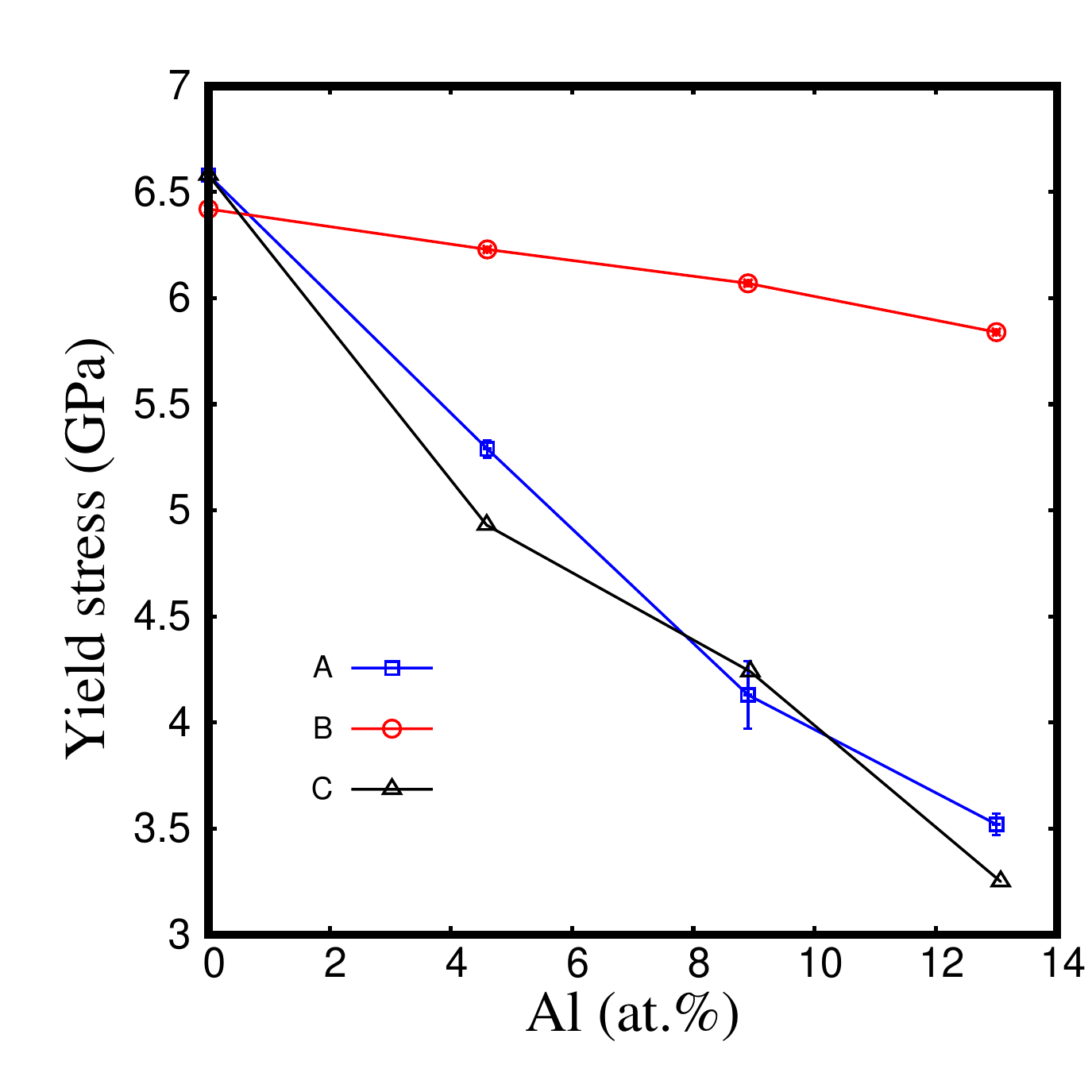} 
\caption{Change in yield stress with Al addition: the plot obtained from MD simulations is indicated by `$A$'; the plot obtained from
a continuum model~\cite{Sylvie} that assumes only homogeneous nucleation is indicated by `$B$'; and the curve $C$ is a combination
of the continuum model for homogeneous nucleation and the fit for heterogeneous nucleation (in terms of the ratio of the unstable
and stable stacking fault energies). See the text for details of $C$.}\label{YS_both}
\end{figure}

Assuming that the dislocation loop is on (111) plane and has $\frac{1}{6}[11\bar{2}]$ Burgers vector, the calculated values of $\sigma_{nuc}$ for Cu and Cu-Al alloys are shown in Figure.~\ref{YS_both} along with the yield stress values obtained from the MD simulations. As seen from the Figure.~\ref{YS_both} and Table.~\ref{bv2} that the tensile stress required for nucleation of partial dislocation decreases with increasing Al.

From the calculation of nucleation stress, we notice that the stress needed for the dislocation nucleation depends on the stacking fault energy. $\sigma_{nuc}$ decreases with decreasing SFE. Since the addition of Al lowers the SFE of Cu-Al alloys, the yield stress  (nucleation stress) reduces with Al addition leading to softening. 

However, as is clear from Figure.~\ref{YS_both} and Table.~\ref{bv2}, even though the continuum model
agrees very well with MD simulations, there are deviations in Cu-Al alloys; the deviations increase with
increasing Al content. The MD simulations show that the yield stress is lower than that predicted
by the continuum model. This, we believe, is because of the ease of heterogeneous nucleation and we discuss this further in the next subsection.
 

\begin{table}[htbp]
\caption{Stress required for nucleation of partial dislocation with Burgers vector $\frac{1}{6}[11\bar{2}]$ in (111) plane
 for loading along $\langle 001 \rangle$ direction; $\sigma_{nuc}$ is calculated from the continuum model and $\sigma_y$ is obtained from MD simulations.}
{\begin{tabular}[l]{@{}lcccccc} \hline

  \footnotesize Material   &   {\footnotesize $ \sigma_{nuc}$ ($GPa$) } & \footnotesize  $\sigma_y$ ($GPa$) &\footnotesize $ \Delta = \sigma_{y}- \sigma_{nuc}$ ($GPa$)\\ \hline

  \footnotesize Pure Cu    & \footnotesize 6.42 $\pm0.0$&\footnotesize 6.58 $\pm0.016$ & \footnotesize -0.16 \\                
  \footnotesize Cu-4.6 at.\%Al     & \footnotesize  6.230 $\pm0.01$ & \footnotesize      5.29$\pm0.04$                   & \footnotesize 0.94\\                                                         
  \footnotesize Cu-8.9 at.\%Al & \footnotesize  6.073 $\pm0.01$ & \footnotesize       4.13$\pm0.16$    & \footnotesize 1.94                 \\                                                         
 \footnotesize  Cu-13.0 at.\%Al    &  \footnotesize 5.841$\pm0.0$ &  \footnotesize       3.52$\pm0.05$   & \footnotesize 2.32 \\                                                        \hline

 \end{tabular}}
 \label{bv2}
\end{table}

\subsection{Solid solution softening}

As indicated in the introduction, solid solution softening has been reported in MD simulations of several copper based systems; specifically,
it has been reported in Cu-Pb system by  Rupert~\cite{TJRupert}, Cu-Ag system by Amigo et al.~\cite{CuAg}, and Cu-Sb system by Rajgarhia et al.~\cite{RKRajgarhia}. 

Rupert reports that in nanocrystalline materials yield strength is proportional to the Young's modulus similar to the metallic glasses, and since the addition of Pb to Cu reduces the Young's modulus, the material shows softening behaviour. But our studies are on single crystals where no sources of dislocations such as grain boundaries exist; so, even though we also see decrease in Young's modulus with addition of Al to Cu, his explanation is not directly applicable for the system under study. 

Rajgarhia et al.~\cite{RKRajgarhia} report that the energy barrier for the Shockley partial dislocation nucleation depends on the ratio  
$\gamma_{sf}/\gamma_{usf}$. Since in Cu-Sb system, the amount of solute added is very small and the SFE change is also small i.e., from 
45.15 to 44.12 mJ/m$^2$ in the range 0 - 2 at.\%Sb. The dislocation nucleation stress is attributed only to local change in unstable stacking 
fault energy near the solute atoms. Similarly in Cu-Ag~\cite{CuAg} system also the reasons for decreasing yield stress are discussed with respect
to the change in unstable stacking fault energy near Ag atoms. In both the systems, the addition of solute decreases the unstable stacking 
fault energy near the solute atoms and  which in turn reduces the stress required for dislocation nucleation. The reduction in energy 
occur only if solute atoms are present at the fault plane and the dislocation nucleation is heterogeneous. 
However, as we have seen, in our simulations, we observe homogeneous nucleation even in systems with 4.6
and 8.9 at.\% Al alloys. Hence, an explanation that is based only on heterogeneous nucleation is not
complete.

Further, as indicated in the introduction, studies in which the SFE of copper and aluminium was changed by using different interatomic potentials, 
as the SFE is decreased, the deformation is dominated by partials~\cite{DiffSFECu,DiffSFEAl}; specifically, in the case of copper bicrystals, the 
same trend, namely decrease in yield stress with decreasing SFE is observed. Thus, our results are consistent with these studies even though
the change in SFE in our study is due to change in compositions and is not due to the use of different interatomic potentials.

In our study, we find that dislocation nucleation occurs both at Al sites and away from Al sites (even in systems with 4.6 and 8.94 at.\% Al 
cases) -- see Fig.~\ref{hetero}; thus, we see both homogeneous and heterogeneous nucleation of dislocations and hence, the softening in our case 
can not be attributed to heterogeneous nucleation alone. In fact, as seen from Figure.~\ref{YS_both} and Table.~\ref{bv2}, even in the alloy with higher amount of Al, the homogeneous nucleation model can explain most of the drop in yield strength as compared to pure copper. This is because, in our studies, the SFE change is significant i.e $\gamma_{sf}$ decreases from 43 to 6 
$mJ/m^2$ in the composition Al range 0 - 13 at.\%, irrespective of the position of Al atoms in the system. Since the addition of Al atoms reduces 
the SFE of the system, ease of partial dislocation nucleation increases. Therefore, the addition of Al leads to solid solution softening.  
Having said that, we do see a discrepancy between the results obtained from a continuum model and those from the simulations. In order to account for this difference, we have fit the difference between the MD simulations and the continuum model to an equation of the form $A \exp{\left[ B \frac{\gamma_{usf}}{\gamma_sf} \right]}$. This fit, which indicates the reduction in yield stress due to the decrease
in barrier for heterogeneous nucleation, can thus account for the reduction in yield stress over
and above that is caused by the changes in SFE, moduli and lattice parameter changes associated with
the alloying addition -- see plot C of Figure.~\ref{YS_both}.

Conventionally, the addition of a solute is expected to increase the yield stress in alloys~\cite{Dieter,Courtney,MayersChawla}. And also 
experimental studies on Cu-Al alloys~\cite{Rohatgi,Yin2016,Liu2015a,Szczerba2017} show solid solution strengthening i.e  the strength of the 
material increases with increasing the Al content. In contrast, in our simulations, we see an anomalous behaviour -- namely ``solid solution 
softening". This, we believe, is because our simulations are carried out on defect free single crystals and hence the deformation is dislocation
 nucleation limited. On the other hand, experiments are typically done on polycrystalline materials containing both  dislocations and grain 
 boundaries; the latter  can act as both source and sink for dislocations. Hence direct quantitative comparison between experiments and our 
 simulation results is not possible. Having said that, our simulation results show that in the case of dislocation nucleation limited plasticity 
 alloying additions can lead to softening due to their reducing the SFE of the material. Thus, studies on small scale pillars and whiskers
 in these materials might be interesting as they have the potential to lead to interesting deformation behaviour.

\section{Conclusions} \label{Section6}

\begin{itemize}
\item Using thermodynamic integration, it is possible to evaluate the stacking fault energies (SFEs) at the temperatures
at which deformation is carried out in MD simulations in metals and alloys; these SFE values, thus obtained, are the best 
suited to analyse and understand deformation behaviour studies carried out using MD simulations with the same interatomic
potentials;

\item In the deformation behaviour of Cu rich Cu-Al alloys, SFE plays a crucial role; with increasing Al content, SFE decreases. Hence,
it is easier for the system to nucleate partial dislocations; therefore, yield occurs at lower applied stress with increasing Al content. 

\item We have shown that the yield stress values obtained from the simulations agree fairly well with the stress required for homogeneous nucleation of partials calculated using a continuum model (and corrected
for ease of heterogeneous nucleation in alloys with increasing Al content). Because of decreasing Young's modulus and yield stress, the system stores less 
elastic energy and hence the dislocation density decreases with Al addition. 

\item The solid solution softening resulting from decrease in SFE in the dislocation nucleation limited plasticity regime indicates
that nano pillar and whisker experiments in these alloys will be interesting from a fundamental point of view.
\end{itemize}

\section*{Acknowledgements}

We thank Saransh Singh, Prachi Limaye, and, Arijit Roy for useful discussions; 
the high performance computing facilities (i) Dendrite and Space-Time, IIT Bombay, 
(ii) Spinode -- the DST-FIST HPC facility,  and (iii) and C-DAC, Pune for the computational resources; 
and, DST, Government of India for funding this project (14DST017). 

\section*{Data availability}

The raw/processed data required to reproduce these findings cannot be shared at this time due to technical or time limitations. 

\section*{Supplementary information}

The scripts used for thermodynamic integration and the stress-strain response
for the PZ potential are available at \verb+http://dx.doi.org/10.17632/vhkhcs4vnt.1+

\bibliographystyle{elsarticle-num-names} 

\bibliography{kamalakshiguruprita}

\begin{thebibliography}{61}
\expandafter\ifx\csname natexlab\endcsname\relax\def\natexlab#1{#1}\fi
\providecommand{\url}[1]{\texttt{#1}}
\providecommand{\href}[2]{#2}
\providecommand{\path}[1]{#1}
\providecommand{\DOIprefix}{doi:}
\providecommand{\ArXivprefix}{arXiv:}
\providecommand{\URLprefix}{URL: }
\providecommand{\Pubmedprefix}{pmid:}
\providecommand{\doi}[1]{\href{http://dx.doi.org/#1}{\path{#1}}}
\providecommand{\Pubmed}[1]{\href{pmid:#1}{\path{#1}}}
\providecommand{\bibinfo}[2]{#2}
\ifx\xfnm\relax \def\xfnm[#1]{\unskip,\space#1}\fi
\bibitem[{Dieter(2016)}]{Dieter}
\bibinfo{author}{G.~E. Dieter}, \bibinfo{title}{Mechanical Metallurgy},
  \bibinfo{edition}{3} ed., \bibinfo{publisher}{McGraw Hill Education (India)
  Private Limited}, \bibinfo{address}{New Delhi}, \bibinfo{year}{2016}.
\bibitem[{Courtney(2000)}]{Courtney}
\bibinfo{author}{T.~H. Courtney}, \bibinfo{title}{Mechanical Behavior of
  Materials}, \bibinfo{edition}{2} ed., \bibinfo{publisher}{McGraw Hill Custom
  Pub.}, \bibinfo{address}{United States of America}, \bibinfo{year}{2000}.
\bibitem[{Meyers and Chawla(2009)}]{MayersChawla}
\bibinfo{author}{M.~A. Meyers}, \bibinfo{author}{K.~K. Chawla},
  \bibinfo{title}{Mechanical Behavior of Materials}, \bibinfo{edition}{2} ed.,
  \bibinfo{publisher}{Cambridge University Press}, \bibinfo{address}{New York},
  \bibinfo{year}{2009}.
\bibitem[{Rupert(2014)}]{TJRupert}
\bibinfo{author}{T.~J. Rupert},
\newblock \bibinfo{title}{{Solid solution strengthening and softening due to
  collective nanocrystalline deformation physics}},
\newblock \bibinfo{journal}{Scripta Mater.} \bibinfo{volume}{81}
  (\bibinfo{year}{2014}) \bibinfo{pages}{44--47}. \URLprefix
  \url{https://doi.org/10.1016/j.scriptamat.2014.03.006}.
\bibitem[{Amigo et~al.(2014)Amigo, Guti{\'{e}}rrez, and Ignat}]{CuAg}
\bibinfo{author}{N.~Amigo}, \bibinfo{author}{G.~Guti{\'{e}}rrez},
  \bibinfo{author}{M.~Ignat},
\newblock \bibinfo{title}{{Atomistic simulation of single crystal copper
  nanowires under tensile stress: Influence of silver impurities in the
  emission of dislocations}},
\newblock \bibinfo{journal}{Comp. Mater. Sci.} \bibinfo{volume}{87}
  (\bibinfo{year}{2014}) \bibinfo{pages}{76--82}. \URLprefix
  \url{https://doi.org/10.1016/j.commatsci.2014.02.014}.
\bibitem[{Rajgarhia et~al.(2009)Rajgarhia, Spearot, and Saxena}]{RKRajgarhia}
\bibinfo{author}{R.~K. Rajgarhia}, \bibinfo{author}{D.~E. Spearot},
  \bibinfo{author}{A.~Saxena},
\newblock \bibinfo{title}{{Heterogeneous dislocation nucleation in single
  crystal copper-antimony solid-solution alloys}},
\newblock \bibinfo{journal}{Modell. Simul. Mater. Sci. Eng.}
  \bibinfo{volume}{17} (\bibinfo{year}{2009}) \bibinfo{pages}{55001--13}.
  \URLprefix \url{https://doi.org/10.1088/0965-0393/17/5/055001}.
\bibitem[{Borovikov et~al.(2016)Borovikov, Mendelev, and King}]{DiffSFECu}
\bibinfo{author}{V.~Borovikov}, \bibinfo{author}{M.~I. Mendelev},
  \bibinfo{author}{A.~H. King},
\newblock \bibinfo{title}{{Effects of stable and unstable stacking fault energy
  on dislocation nucleation in nano-crystalline metals}},
\newblock \bibinfo{journal}{Modell. Simul. Mater. Sci. Eng.}
  \bibinfo{volume}{24} (\bibinfo{year}{2016}) \bibinfo{pages}{85017--14}.
  \URLprefix \url{https://doi.org/10.1088/0965-0393/24/8/085017}.
\bibitem[{Shimokawa et~al.(2004)Shimokawa, Nakatani, and Kitagawa}]{DiffSFEAl}
\bibinfo{author}{T.~Shimokawa}, \bibinfo{author}{A.~Nakatani},
  \bibinfo{author}{H.~Kitagawa},
\newblock \bibinfo{title}{{Mechanical Properties Depending on Grain Sizes of
  Face-Centered-Cubic Nanocrystalline Metals Using Molecular Dynamics
  Simulation (Investigation of Stacking Fault Energy's Influence)}},
\newblock \bibinfo{journal}{JSME Int. J.} \bibinfo{volume}{47}
  (\bibinfo{year}{2004}) \bibinfo{pages}{1708--1715}. \URLprefix
  \url{https://doi.org/10.1299/jsmea.47.83}.
\bibitem[{Rohatgi et~al.(2001)Rohatgi, Vecchio, and Gray}]{Rohatgi}
\bibinfo{author}{A.~Rohatgi}, \bibinfo{author}{K.~S. Vecchio},
  \bibinfo{author}{G.~T. Gray},
\newblock \bibinfo{title}{{The influence of stacking fault energy on the
  mechanical behavior of Cu and Cu-Al alloys: Deformation twinning, work
  hardening, and dynamic recovery}},
\newblock \bibinfo{journal}{Metall. Mater. Trans. A} \bibinfo{volume}{32}
  (\bibinfo{year}{2001}) \bibinfo{pages}{135--145}. \URLprefix
  \url{https://doi.org/10.1007/s11661-001-0109-7}.
\bibitem[{Carter and Ray(1977)}]{CarterRay}
\bibinfo{author}{C.~B. Carter}, \bibinfo{author}{I.~L.~F. Ray},
\newblock \bibinfo{title}{{On the stacking-fault energies of copper alloys}},
\newblock \bibinfo{journal}{Philos. Mag.} \bibinfo{volume}{35}
  (\bibinfo{year}{1977}) \bibinfo{pages}{189--200}. \URLprefix
  \url{https://doi.org/10.1080/14786437708235982}.
\bibitem[{Lu et~al.(2016)Lu, Chen, and Meng}]{phasedia}
\bibinfo{author}{B.~Lu}, \bibinfo{author}{K.~Chen}, \bibinfo{author}{W.~J.
  Meng},
\newblock \bibinfo{title}{{Quantification of Thermal Resistance of
  Transient-Liquid- Phase Bonded Cu / Al / Cu Interfaces for Assembly of
  Cu-Based Microchannel Heat Exchangers}},
\newblock \bibinfo{journal}{J. Micro Nano-Manuf.} \bibinfo{volume}{1}
  (\bibinfo{year}{2016}) \bibinfo{pages}{1--10}. \URLprefix
  \url{https://doi.org/10.1115/1.4024683}.
\bibitem[{Yin et~al.(2016)Yin, Sun, Yang, Gong, and Zhu}]{Yin2016}
\bibinfo{author}{Z.~Yin}, \bibinfo{author}{L.~Sun}, \bibinfo{author}{J.~Yang},
  \bibinfo{author}{Y.~Gong}, \bibinfo{author}{X.~Zhu},
\newblock \bibinfo{title}{{Mechanical behavior and deformation kinetics of
  gradient structured Cu-Al alloys with varying stacking fault energy}},
\newblock \bibinfo{journal}{J. Alloy. Compd.} \bibinfo{volume}{687}
  (\bibinfo{year}{2016}) \bibinfo{pages}{152--160}. \URLprefix
  \url{https://doi.org/10.1016/j.jallcom.2016.06.155}.
\bibitem[{Liu et~al.(2015)Liu, Zhang, Li, An, and Zhang}]{Liu2015a}
\bibinfo{author}{R.~Liu}, \bibinfo{author}{Z.~J. Zhang}, \bibinfo{author}{L.~L.
  Li}, \bibinfo{author}{X.~H. An}, \bibinfo{author}{Z.~F. Zhang},
\newblock \bibinfo{title}{{Microscopic mechanisms contributing to the
  synchronous improvement of strength and plasticity (SISP) for TWIP copper
  alloys}},
\newblock \bibinfo{journal}{Sci. Rep.} \bibinfo{volume}{5}
  (\bibinfo{year}{2015}) \bibinfo{pages}{1--7}. \URLprefix
  \url{https://doi.org/10.1038/srep09550}.
\bibitem[{Szczerba and Szczerba(2017)}]{Szczerba2017}
\bibinfo{author}{M.~J. Szczerba}, \bibinfo{author}{M.~S. Szczerba},
\newblock \bibinfo{title}{{Slip versus twinning in low and very low
  stacking-fault energy Cu-Al alloy single crystals}},
\newblock \bibinfo{journal}{Acta Mater.} \bibinfo{volume}{133}
  (\bibinfo{year}{2017}) \bibinfo{pages}{109--119}. \URLprefix
  \url{https://doi.org/10.1016/j.actamat.2017.05.011}.
\bibitem[{Bei et~al.(2007)Bei, Shim, George, Miller, Herbert, and
  Pharr}]{Bei2007}
\bibinfo{author}{H.~Bei}, \bibinfo{author}{S.~Shim}, \bibinfo{author}{E.~P.
  George}, \bibinfo{author}{M.~K. Miller}, \bibinfo{author}{E.~Herbert},
  \bibinfo{author}{G.~M. Pharr},
\newblock \bibinfo{title}{Compressive strengths of molybdenum alloy
  micro-pillars prepared using a new technique},
\newblock \bibinfo{journal}{Scripta Mater.} \bibinfo{volume}{57}
  (\bibinfo{year}{2007}) \bibinfo{pages}{397--400}. \URLprefix
  \url{https://doi.org/10.1016/j.scriptamat.2007.05.010}.
\bibitem[{Lu et~al.(2011)Lu, Song, Huang, and Lou}]{lu2011surface}
\bibinfo{author}{Y.~Lu}, \bibinfo{author}{J.~Song}, \bibinfo{author}{J.~Y.
  Huang}, \bibinfo{author}{J.~Lou},
\newblock \bibinfo{title}{Surface dislocation nucleation mediated deformation
  and ultrahigh strength in sub-10-nm gold nanowires},
\newblock \bibinfo{journal}{Nano Res.} \bibinfo{volume}{4}
  (\bibinfo{year}{2011}) \bibinfo{pages}{1261--1267}. \URLprefix
  \url{https://doi.org/10.1007/s12274-011-0177-y}.
\bibitem[{Richter et~al.(2009)Richter, Hillerich, Gianola, Mönig, Kraft, and
  Volkert}]{Gunther2009}
\bibinfo{author}{G.~Richter}, \bibinfo{author}{K.~Hillerich},
  \bibinfo{author}{D.~S. Gianola}, \bibinfo{author}{R.~Mönig},
  \bibinfo{author}{O.~Kraft}, \bibinfo{author}{C.~A. Volkert},
\newblock \bibinfo{title}{Ultrahigh strength single crystalline nanowhiskers
  grown by physical vapor deposition},
\newblock \bibinfo{journal}{Nano Lett.} \bibinfo{volume}{9}
  (\bibinfo{year}{2009}) \bibinfo{pages}{3048--3052}. \URLprefix
  \url{https://doi.org/10.1021/nl9015107}.
\bibitem[{Chen et~al.(2015)Chen, He, Shin, Richter, and Gianola}]{Chen2015}
\bibinfo{author}{L.~Y. Chen}, \bibinfo{author}{M.-r. He},
  \bibinfo{author}{J.~Shin}, \bibinfo{author}{G.~Richter},
  \bibinfo{author}{D.~S. Gianola},
\newblock \bibinfo{title}{Measuring surface dislocation nucleation in
  defect-scarce nanostructures},
\newblock \bibinfo{journal}{Nature mater.} \bibinfo{volume}{14}
  (\bibinfo{year}{2015}) \bibinfo{pages}{707--713}. \URLprefix
  \url{https://doi.org/10.1038/nmat4288}.
\bibitem[{{S. Plimpton}(1995)}]{lammps}
\bibinfo{author}{{S. Plimpton}},
\newblock \bibinfo{title}{{Fast Parallel Algorithms for Short-Range Molecular
  Dynamics}},
\newblock \bibinfo{journal}{J. Comput. Phys.} \bibinfo{volume}{117}
  (\bibinfo{year}{1995}) \bibinfo{pages}{1--19}. \URLprefix
  \url{https://doi.org/10.1006/jcph.1995.1039}.
\bibitem[{Stukowski(2009)}]{ovito}
\bibinfo{author}{A.~Stukowski},
\newblock \bibinfo{title}{{Visualization and analysis of atomistic simulation
  data with OVITO – the Open Visualization Tool}},
\newblock \bibinfo{journal}{Modelling Simul. Mater. Sci. Eng.}
  \bibinfo{volume}{18} (\bibinfo{year}{2009}) \bibinfo{pages}{015012--7}.
  \URLprefix \url{https://doi.org/10.1088/0965-0393/18/1/015012}.
\bibitem[{Becker et~al.(2013)Becker, Tavazza, Trautt, and {Buarque De
  Macedo}}]{NISTPotential}
\bibinfo{author}{C.~A. Becker}, \bibinfo{author}{F.~Tavazza},
  \bibinfo{author}{Z.~T. Trautt}, \bibinfo{author}{R.~A. {Buarque De Macedo}},
\newblock \bibinfo{title}{{Considerations for choosing and using force fields
  and interatomic potentials in materials science and engineering}},
\newblock \bibinfo{journal}{Curr. Opin. Solid State Mater. Sci.}
  \bibinfo{volume}{17} (\bibinfo{year}{2013}) \bibinfo{pages}{277--283}.
  \URLprefix \url{https://doi.org/10.1016/j.cossms.2013.10.001}.
\bibitem[{Zhou et~al.(2004)Zhou, Johnson, and Wadley}]{Zhou2004}
\bibinfo{author}{X.~W. Zhou}, \bibinfo{author}{R.~A. Johnson},
  \bibinfo{author}{H.~N.~G. Wadley},
\newblock \bibinfo{title}{{Misfit-energy-increasing dislocations in
  vapor-deposited CoFe/NiFe multilayers}},
\newblock \bibinfo{journal}{Phys. Rev. B - Condensed Matter and Materials
  Physics} \bibinfo{volume}{69} (\bibinfo{year}{2004}) \bibinfo{pages}{144113}.
  \URLprefix \url{https://doi.org/10.1103/PhysRevB.69.144113}.
\bibitem[{Ward et~al.(2012)Ward, Agrawal, Flores, and Windl}]{ZW}
\bibinfo{author}{L.~Ward}, \bibinfo{author}{A.~Agrawal}, \bibinfo{author}{K.~M.
  Flores}, \bibinfo{author}{W.~Windl},
\newblock \bibinfo{title}{{Rapid Production of Accurate Embedded Atom Method
  Potentials for Metal Alloys}},
\newblock \bibinfo{journal}{arXiv:1209.0619}  (\bibinfo{year}{2012}).
  \URLprefix \url{https://arxiv.org/abs/1209.0619}.
\bibitem[{Cain and Thomas(1971)}]{CainThomas}
\bibinfo{author}{L.~S. Cain}, \bibinfo{author}{J.~Thomas},
\newblock \bibinfo{title}{{Elastic Constants of $\ensuremath{\alpha}$-Phase
  Cu-Al Alloys}},
\newblock \bibinfo{journal}{Phys. Rev. B} \bibinfo{volume}{4}
  (\bibinfo{year}{1971}) \bibinfo{pages}{4245--4255}. \URLprefix
  \url{https://doi.org/10.1103/PhysRevB.4.4245}.
\bibitem[{Seshadri and Downie(1979)}]{Seshadri1979}
\bibinfo{author}{S.~K. Seshadri}, \bibinfo{author}{D.~B. Downie},
\newblock \bibinfo{title}{High-temperature latticeparan,eters of
  copper-aluminium alloys},
\newblock \bibinfo{journal}{Metal Science} \bibinfo{volume}{13}
  (\bibinfo{year}{1979}) \bibinfo{pages}{696--698}. \URLprefix
  \url{https://doi.org/10.1179/030634579790434330}.
\bibitem[{Tomokiyo et~al.(1987)Tomokiyo, Matsumura, and Kuwano}]{Tomokiyo1987}
\bibinfo{author}{Y.~Tomokiyo}, \bibinfo{author}{S.~Matsumura},
  \bibinfo{author}{N.~Kuwano},
\newblock \bibinfo{title}{Application of higher order laue zone patterns to
  lattice parameter determination in cu-based alloys},
\newblock \bibinfo{journal}{J. Electron Microsc.} \bibinfo{volume}{35}
  (\bibinfo{year}{1987}) \bibinfo{pages}{359--364}. \URLprefix
  \url{https://doi.org/10.1093/oxfordjournals.jmicro.a050589}.
\bibitem[{J.F.Nye(1985)}]{JFNye}
\bibinfo{author}{J.F.Nye}, \bibinfo{title}{{Physical properties of crystals :
  their representations by tensors and matrices}}, \bibinfo{edition}{2} ed.,
  \bibinfo{publisher}{Oxford University Press}, \bibinfo{year}{1985}.
\bibitem[{Thompson(line)}]{EConstant}
\bibinfo{author}{A.~Thompson}, \bibinfo{title}{Lammps molecular dynamics
  simulator}, \bibinfo{howpublished}{LAMMPS}, \bibinfo{year}{2020. [Online]}.
  \URLprefix \url{https://lammps.sandia.gov/doc/Examples.html}.
\bibitem[{Moment(1972)}]{Moment1972}
\bibinfo{author}{R.~L. Moment},
\newblock \bibinfo{title}{{Elastic stiffnesses of copper-tin and
  copper-aluminum alloy single crystals}},
\newblock \bibinfo{journal}{J. Appl. Phys.} \bibinfo{volume}{43}
  (\bibinfo{year}{1972}) \bibinfo{pages}{4419--4424}. \URLprefix
  \url{https://doi.org/10.1063/1.1660937}. \DOIprefix\doi{10.1063/1.1660937}.
\bibitem[{Neighbours and Smiths(1954)}]{Neighbourst}
\bibinfo{author}{J.~R. Neighbours}, \bibinfo{author}{C.~S. Smiths},
\newblock \bibinfo{title}{{The elastic constants of copper alloys}},
\newblock \bibinfo{journal}{Acta metallurgica} \bibinfo{volume}{2}
  (\bibinfo{year}{1954}) \bibinfo{pages}{591--596}. \URLprefix
  \url{https://doi.org/10.1016/0001-6160(54)90193-5}.
\bibitem[{Wang and Wang(2009)}]{Yun-Jiang2009}
\bibinfo{author}{Y.~J. Wang}, \bibinfo{author}{C.~Y. Wang},
\newblock \bibinfo{title}{Influence of alloying elements on the elastic
  properties of ternary and quaternary nickel-base superalloys},
\newblock \bibinfo{journal}{Philos. Mag.} \bibinfo{volume}{89}
  (\bibinfo{year}{2009}) \bibinfo{pages}{2935--2947}. \URLprefix
  \url{https://doi.org/10.1080/14786430903140747}.
\bibitem[{S\"oderlind et~al.(1993)S\"oderlind, Eriksson, Wills, and
  Boring}]{PhysRevB1993}
\bibinfo{author}{P.~S\"oderlind}, \bibinfo{author}{O.~Eriksson},
  \bibinfo{author}{J.~M. Wills}, \bibinfo{author}{A.~M. Boring},
\newblock \bibinfo{title}{Theory of elastic constants of cubic transition
  metals and alloys},
\newblock \bibinfo{journal}{Phys. Rev. B} \bibinfo{volume}{48}
  (\bibinfo{year}{1993}) \bibinfo{pages}{5844--5851}. \URLprefix
  \url{https://link.aps.org/doi/10.1103/PhysRevB.48.5844}.
\bibitem[{Wu et~al.(2008)Wu, Hu, and Han}]{WU2008}
\bibinfo{author}{Y.~Wu}, \bibinfo{author}{W.~Hu}, \bibinfo{author}{S.~Han},
\newblock \bibinfo{title}{First-principles calculation of the elastic
  constants, the electronic density of states and the ductility mechanism of
  the intermetallic compounds: Yag, ycu and yrh},
\newblock \bibinfo{journal}{Physica B} \bibinfo{volume}{403}
  (\bibinfo{year}{2008}) \bibinfo{pages}{3792--3797}. \URLprefix
  \url{https://doi.org/10.1016/j.physb.2008.07.009}.
\bibitem[{Ezaz et~al.(2011)Ezaz, Sangid, and Sehitoglu}]{Ezaz2011}
\bibinfo{author}{T.~Ezaz}, \bibinfo{author}{M.~D. Sangid},
  \bibinfo{author}{H.~Sehitoglu},
\newblock \bibinfo{title}{{Energy barriers associated with slip–twin
  interactions}},
\newblock \bibinfo{journal}{Phil. Mag.} \bibinfo{volume}{91}
  (\bibinfo{year}{2011}) \bibinfo{pages}{1464--1488}. \URLprefix
  \url{https://doi.org/10.1080/14786435.2010.541166}.
\bibitem[{Liang et~al.(2015)Liang, L{\"u}, Tieu, Xing, Lin-Qing, and
  Michal}]{Lu2015}
\bibinfo{author}{Z.~Liang}, \bibinfo{author}{C.~L{\"u}},
  \bibinfo{author}{K.~Tieu}, \bibinfo{author}{Z.~Xing},
  \bibinfo{author}{P.~Lin-Qing}, \bibinfo{author}{G.~Michal},
\newblock \bibinfo{title}{{Molecular dynamics simulation on generalized
  stacking fault energies of FCC metals under preloading stress}},
\newblock \bibinfo{journal}{Chinese Phys. B} \bibinfo{volume}{24}
  (\bibinfo{year}{2015}) \bibinfo{pages}{088106}. \URLprefix
  \url{https://doi.org/10.1088/1674-1056/24/8/088106}.
\bibitem[{Zimmerman et~al.(2000)Zimmerman, Gao, and Abraham}]{Zimmerman2000}
\bibinfo{author}{J.~A. Zimmerman}, \bibinfo{author}{H.~Gao},
  \bibinfo{author}{F.~F. Abraham},
\newblock \bibinfo{title}{{Generalized stacking fault energies for embedded
  atom FCC metals}},
\newblock \bibinfo{journal}{Modelling Simul. Mater. Sci.} \bibinfo{volume}{8}
  (\bibinfo{year}{2000}) \bibinfo{pages}{103–115}. \URLprefix
  \url{https://doi.org/10.1088/0965-0393/8/2/302}.
\bibitem[{Chassagne et~al.(2011)Chassagne, Legros, and Rodney}]{Chassagne2011}
\bibinfo{author}{M.~Chassagne}, \bibinfo{author}{M.~Legros},
  \bibinfo{author}{D.~Rodney},
\newblock \bibinfo{title}{{Atomic-scale simulation of screw
  dislocation/coherent twin boundary interaction in Al, Au, Cu and Ni}},
\newblock \bibinfo{journal}{Acta Mater.} \bibinfo{volume}{59}
  (\bibinfo{year}{2011}) \bibinfo{pages}{1456--1463}. \URLprefix
  \url{https://doi.org/10.1016/j.actamat.2010.11.007}.
\bibitem[{Swygenhoven and Derlet(2004)}]{Swygenhoven2004}
\bibinfo{author}{H.~V. A.~N. Swygenhoven}, \bibinfo{author}{P.~M. Derlet},
\newblock \bibinfo{title}{{Stacking fault energies and slip in nanocrystalline
  metals}},
\newblock \bibinfo{journal}{Nat. Mater.} \bibinfo{volume}{3}
  (\bibinfo{year}{2004}) \bibinfo{pages}{399--403}. \URLprefix
  \url{https://doi.org/10.1038/nmat1136}.
\bibitem[{Smit and Frenkel(2002)}]{FrenkelSmit}
\bibinfo{author}{B.~Smit}, \bibinfo{author}{D.~Frenkel},
  \bibinfo{title}{Understanding Molecular Simulation: From Algorithms to
  Applications}, \bibinfo{edition}{2} ed., \bibinfo{publisher}{Academic Press},
  \bibinfo{address}{San Diego}, \bibinfo{year}{2002}.
\bibitem[{Freitas et~al.(2016)Freitas, Asta, and {De Koning}}]{Freitas2016}
\bibinfo{author}{R.~Freitas}, \bibinfo{author}{M.~Asta},
  \bibinfo{author}{M.~{De Koning}},
\newblock \bibinfo{title}{{Nonequilibrium free-energy calculation of solids
  using LAMMPS}},
\newblock \bibinfo{journal}{Comp. Mater. Sci.} \bibinfo{volume}{112}
  (\bibinfo{year}{2016}) \bibinfo{pages}{333--341}. \URLprefix
  \url{http://dx.doi.org/10.1016/j.commatsci.2015.10.050}.
\bibitem[{Lenkkeri and Lahteenkorva(1973)}]{Lenkkeri1973}
\bibinfo{author}{J.~T. Lenkkeri}, \bibinfo{author}{E.~E. Lahteenkorva},
\newblock \bibinfo{title}{{An investigation of elastic moduli of alpha
  copper-aluminium alloys}},
\newblock \bibinfo{journal}{J. Phys. F Met. Phys.} \bibinfo{volume}{3}
  (\bibinfo{year}{1973}) \bibinfo{pages}{1781--1788}. \URLprefix
  \url{https://doi.org/10.1088/0305-4608/3/10/013}.
\bibitem[{Koster and Walter(1951)}]{Koster}
\bibinfo{author}{W.~Koster}, \bibinfo{author}{R.~Walter},
\newblock \bibinfo{title}{{Relation between the modulus of elasticity of binary
  alloys and thier structure}},
\newblock \bibinfo{journal}{National Advisory Committe for Aeuronautics}
  (\bibinfo{year}{1951}).
\bibitem[{Salehinia and Bahr(2014)}]{Salehinia2014}
\bibinfo{author}{I.~Salehinia}, \bibinfo{author}{D.~F. Bahr},
\newblock \bibinfo{title}{{Crystal orientation effect on dislocation nucleation
  and multiplication in FCC single crystal under uniaxial loading}},
\newblock \bibinfo{journal}{Int. J. Plasticity} \bibinfo{volume}{52}
  (\bibinfo{year}{2014}) \bibinfo{pages}{133--146}. \URLprefix
  \url{https://doi.org/10.1016/j.ijplas.2013.04.010}.
\bibitem[{Zhao et~al.(2009)Zhao, Chen, Shen, and Lu}]{Zhao2009}
\bibinfo{author}{K.~J. Zhao}, \bibinfo{author}{C.~Q. Chen},
  \bibinfo{author}{Y.~P. Shen}, \bibinfo{author}{T.~J. Lu},
\newblock \bibinfo{title}{{Molecular dynamics study on the nano-void growth in
  face-centered cubic single crystal copper}},
\newblock \bibinfo{journal}{Comp. Mater. Sci.} \bibinfo{volume}{46}
  (\bibinfo{year}{2009}) \bibinfo{pages}{749--754}. \URLprefix
  \url{http://dx.doi.org/10.1016/j.commatsci.2009.04.034}.
\bibitem[{Tschopp and McDowell(2008)}]{Tschopp2008}
\bibinfo{author}{M.~A. Tschopp}, \bibinfo{author}{D.~L. McDowell},
\newblock \bibinfo{title}{{Influence of single crystal orientation on
  homogeneous dislocation nucleation under uniaxial loading}},
\newblock \bibinfo{journal}{J. Mech. Phys. Solids} \bibinfo{volume}{56}
  (\bibinfo{year}{2008}) \bibinfo{pages}{1806--1830}. \URLprefix
  \url{https://doi.org/10.1016/j.jmps.2007.11.012}.
  \DOIprefix\doi{10.1016/j.jmps.2007.11.012}.
\bibitem[{Setoodeh et~al.(2008)Setoodeh, Attariani, and
  Khosrownejad}]{Setoodeh2008}
\bibinfo{author}{A.~R. Setoodeh}, \bibinfo{author}{H.~Attariani},
  \bibinfo{author}{M.~Khosrownejad},
\newblock \bibinfo{title}{{Nickel nanowires under uniaxial loads: A molecular
  dynamics simulation study}},
\newblock \bibinfo{journal}{Comp. Mater. Sci.} \bibinfo{volume}{44}
  (\bibinfo{year}{2008}) \bibinfo{pages}{378--384}. \URLprefix
  \url{https://doi.org/10.1016/j.commatsci.2008.03.035}.
\bibitem[{Spearot et~al.(2009)Spearot, Tschopp, and McDowell}]{Spearot2009}
\bibinfo{author}{D.~E. Spearot}, \bibinfo{author}{M.~A. Tschopp},
  \bibinfo{author}{D.~L. McDowell},
\newblock \bibinfo{title}{{Orientation and rate dependence of dislocation
  nucleation stress computed using molecular dynamics}},
\newblock \bibinfo{journal}{Scripta Mater.} \bibinfo{volume}{60}
  (\bibinfo{year}{2009}) \bibinfo{pages}{675--678}. \URLprefix
  \url{https://doi.org/10.1016/j.scriptamat.2008.12.037}.
\bibitem[{Yang et~al.(2009)Yang, Lu, and Zhao}]{Yang2009}
\bibinfo{author}{Z.~Yang}, \bibinfo{author}{Z.~Lu}, \bibinfo{author}{Y.~P.
  Zhao},
\newblock \bibinfo{title}{{Atomistic simulation on size-dependent yield
  strength and defects evolution of metal nanowires}},
\newblock \bibinfo{journal}{Comp. Mater. Sci.} \bibinfo{volume}{46}
  (\bibinfo{year}{2009}) \bibinfo{pages}{142--150}. \URLprefix
  \url{https://doi.org/10.1016/j.commatsci.2009.02.015}.
\bibitem[{Sainath et~al.(2017)Sainath, Rohith, and Choudhary}]{Sainath2017}
\bibinfo{author}{G.~Sainath}, \bibinfo{author}{P.~Rohith},
  \bibinfo{author}{B.~K. Choudhary},
\newblock \bibinfo{title}{Size dependent deformation behaviour and dislocation
  mechanisms in 1 0 0 cu nanowires},
\newblock \bibinfo{journal}{Philos. Mag.} \bibinfo{volume}{97}
  (\bibinfo{year}{2017}) \bibinfo{pages}{2632--2657}. \URLprefix
  \url{https://doi.org/10.1080/14786435.2017.1347300}.
\bibitem[{Farkas and Patrick(2009)}]{Farkas2009}
\bibinfo{author}{D.~Farkas}, \bibinfo{author}{L.~Patrick},
\newblock \bibinfo{title}{{Tensile deformation of fcc Ni as described by an EAM
  potential}},
\newblock \bibinfo{journal}{Philos. Mag.} \bibinfo{volume}{89}
  (\bibinfo{year}{2009}) \bibinfo{pages}{3435--3450}. \URLprefix
  \url{https://doi.org/10.1080/14786430903299329}.
\bibitem[{An et~al.(2009)An, Lin, Qu, Yang, Wu, and Zhang}]{An2009}
\bibinfo{author}{X.~An}, \bibinfo{author}{Q.~Lin}, \bibinfo{author}{S.~Qu},
  \bibinfo{author}{G.~Yang}, \bibinfo{author}{S.~Wu}, \bibinfo{author}{Z.-F.
  Zhang},
\newblock \bibinfo{title}{{Influence of stacking-fault energy on the
  accommodation of severe shear strain in Cu-Al alloys during equal-channel
  angular pressing}},
\newblock \bibinfo{journal}{J. Mater. Res.} \bibinfo{volume}{24}
  (\bibinfo{year}{2009}) \bibinfo{pages}{3636--3646}. \URLprefix
  \url{https://doi.org/10.1557/jmr.2009.0426}.
\bibitem[{Gubicza et~al.(2007)Gubicza, Chinh, and Csan}]{Gubicza2007}
\bibinfo{author}{J.~Gubicza}, \bibinfo{author}{N.~Q. Chinh},
  \bibinfo{author}{T.~Csan},
\newblock \bibinfo{title}{{Microstructure and strength of severely deformed fcc
  metals}},
\newblock \bibinfo{journal}{Mater. Sci. Eng. A.} \bibinfo{volume}{462}
  (\bibinfo{year}{2007}) \bibinfo{pages}{86--90}. \URLprefix
  \url{https://doi.org/10.1016/j.msea.2006.02.455}.
\bibitem[{Rifai et~al.(2018)Rifai, Bagherpour, Yamamoto, Yuasa, and
  Miyamoto}]{Rifai2018}
\bibinfo{author}{M.~Rifai}, \bibinfo{author}{E.~Bagherpour},
  \bibinfo{author}{G.~Yamamoto}, \bibinfo{author}{M.~Yuasa},
  \bibinfo{author}{H.~Miyamoto},
\newblock \bibinfo{title}{{Transition of Dislocation Structures in Severe
  Plastic Deformation and Its Effect on Dissolution in Dislocation Etchant}},
\newblock \bibinfo{journal}{Adv. Mater. Sci. Eng.} \bibinfo{volume}{2018}
  (\bibinfo{year}{2018}) \bibinfo{pages}{1--7}. \URLprefix
  \url{https://doi.org/10.1155/2018/4254156}.
\bibitem[{Rohatgi and Vecchio(2002)}]{Rohatgi2002}
\bibinfo{author}{A.~Rohatgi}, \bibinfo{author}{K.~S. Vecchio},
\newblock \bibinfo{title}{{The variation of dislocation density as a function
  of the stacking fault energy in shock-deformed FCC materials}},
\newblock \bibinfo{journal}{Mater. Sci. Eng. A} \bibinfo{volume}{328}
  (\bibinfo{year}{2002}) \bibinfo{pages}{256--266}. \URLprefix
  \url{https://doi.org/10.1016/S0921-5093(01)01702-6}.
\bibitem[{Rohatgi et~al.(2001)Rohatgi, Vecchio, and {Gray, III}}]{Rohatgi2001a}
\bibinfo{author}{A.~Rohatgi}, \bibinfo{author}{K.~Vecchio},
  \bibinfo{author}{G.~{Gray, III}},
\newblock \bibinfo{title}{{A Metallographic and Quantitative Analysis of the
  Influence of Stacking Fault Energy on Shock Hardening in Cu and Cu – Al
  Alloys}},
\newblock \bibinfo{journal}{Acta Mater.} \bibinfo{volume}{49}
  (\bibinfo{year}{2001}) \bibinfo{pages}{427--438}. \URLprefix
  \url{https://doi.org/10.1016/S1359-6454(00)00335-9}.
\bibitem[{Tao et~al.(2013)Tao, Yang, Xiong, Wu, Zhu, and Wen}]{Tao2013}
\bibinfo{author}{J.~Tao}, \bibinfo{author}{K.~Yang},
  \bibinfo{author}{H.~Xiong}, \bibinfo{author}{X.~Wu},
  \bibinfo{author}{X.~Zhu}, \bibinfo{author}{C.~Wen},
\newblock \bibinfo{title}{{The defect structures and mechanical properties of
  Cu and Cu-Al alloys processed by split Hopkinson pressure bar}},
\newblock \bibinfo{journal}{Mater. Sci. Eng. A} \bibinfo{volume}{580}
  (\bibinfo{year}{2013}) \bibinfo{pages}{406--409}. \URLprefix
  \url{http://dx.doi.org/10.1016/j.msea.2013.05.067}.
\bibitem[{Ye et~al.(2019)Ye, Wu, Liu, Xu, and Li}]{Ye2019}
\bibinfo{author}{T.~Ye}, \bibinfo{author}{Y.~Wu}, \bibinfo{author}{A.~Liu},
  \bibinfo{author}{C.~Xu}, \bibinfo{author}{L.~Li},
\newblock \bibinfo{title}{{Mechanical property and microstructure evolution of
  aged 6063 aluminum alloy under high strain rate deformation}},
\newblock \bibinfo{journal}{Vacuum} \bibinfo{volume}{159}
  (\bibinfo{year}{2019}) \bibinfo{pages}{37--44}. \URLprefix
  \url{https://doi.org/10.1016/j.vacuum.2018.10.013}.
\bibitem[{Aubry et~al.(2011)Aubry, Kang, Ryu, and Cai}]{Sylvie}
\bibinfo{author}{S.~Aubry}, \bibinfo{author}{K.~Kang},
  \bibinfo{author}{S.~Ryu}, \bibinfo{author}{W.~Cai},
\newblock \bibinfo{title}{{Energy barrier for homogeneous dislocation
  nucleation: Comparing atomistic and continuum models}},
\newblock \bibinfo{journal}{Scripta Mater.} \bibinfo{volume}{64}
  (\bibinfo{year}{2011}) \bibinfo{pages}{1043--1046}. \URLprefix
  \url{https://doi.org/10.1016/j.scriptamat.2011.02.023}.
\bibitem[{Hull and Bacon(1988)}]{HullBacon}
\bibinfo{author}{D.~Hull}, \bibinfo{author}{D.~J. Bacon},
  \bibinfo{title}{Introduction to Dislocations}, \bibinfo{edition}{5} ed.,
  \bibinfo{publisher}{Elsevier Limited}, \bibinfo{year}{1988}.
\bibitem[{Weertman and Weertman(1969)}]{WandW}
\bibinfo{author}{J.~Weertman}, \bibinfo{author}{J.~R. Weertman},
  \bibinfo{title}{Elementary Dislocation Theory}, \bibinfo{edition}{5} ed.,
  \bibinfo{publisher}{The MacMillan Company Collier MacMillan Limited},
  \bibinfo{address}{London}, \bibinfo{year}{1969}.
\bibitem[{Warner and Curtin(2009)}]{Warner2009}
\bibinfo{author}{D.~H. Warner}, \bibinfo{author}{W.~A. Curtin},
\newblock \bibinfo{title}{{Origins and implications of temperature-dependent
  activation energy barriers for dislocation nucleation in face-centered cubic
  metals}},
\newblock \bibinfo{journal}{Acta Mater.} \bibinfo{volume}{57}
  (\bibinfo{year}{2009}) \bibinfo{pages}{4267--4277}. \URLprefix
  \url{http://dx.doi.org/10.1016/j.actamat.2009.05.024}.

\end{thebibliography}

\end{document}